\documentclass[preprint,superscriptaddress,preprintnumbers]{revtex4-1}[12pt]
\pdfoutput=1
\usepackage {amsmath, amssymb, graphicx, multirow, rotating, slashed, color, xcolor}
\usepackage{hyperref}

\makeatletter
\def\@seccntformat#1{\csname the#1\endcsname.\quad} 

\renewcommand\section{\@startsection
  {section}{1}{0mm}
  {-\baselineskip}
  {0.5\baselineskip}
  {\normalfont\normalsize\bf}}

\renewcommand\subsection{\@startsection
  {subsection}{2}{0mm}
  {-\baselineskip}
  {0.5\baselineskip}
  {\normalfont\normalsize\it}}
\makeatother

\newcommand{\unit} [1] {\; \mathrm {#1}}

\newcommand{\cO}{\mathcal{O}}

\newcommand{\W}{\mathcal{W}}
\newcommand{\Wdark}{\mathcal{W}_{d}}

\newcommand{\wt}{\widetilde}

\newcommand{\mdark}{m_{\gamma_d}}

\newcommand{\mgaugino}{ m_{\tilde \gamma_d}}
\newcommand{\Udark}{U(1)_{d}}
\newcommand{\dgauge}{\gamma_d}
\newcommand{\dgaugino}{\tilde \gamma_d}
\newcommand{\gdark}{g_{d}}

 \DeclareMathOperator{\kev}{keV} \DeclareMathOperator{\mev}{MeV} \DeclareMathOperator{\gev}{GeV} \DeclareMathOperator{\tev}{TeV}    \DeclareMathOperator{\s}{sec}    \DeclareMathOperator{\few}{few}

\newcommand{\pL}{\left(} \newcommand{\pR}{\right)} \newcommand{\bL}{\left[} \newcommand{\bR}{\right]} \newcommand{\cbL}{\left\{}  \newcommand{\mL}{\left|} \newcommand{\mR}{\right|}

\newcommand{\half}{\frac12}
\newcommand{\beq}{\begin{equation}} \newcommand{\eeq}{\end{equation}}
\newcommand{\bea}{\begin{eqnarray}} \newcommand{\eea}{\end{eqnarray}}
\newcommand{\alg}[1]{\begin{align} \begin{split} #1 \end{split}  \end{align}}

\newcommand{\Eq}[1]{Eq.~(\ref{#1})} \newcommand{\Eqs}[2]{Eqs.~(\ref{#1}) and (\ref{#2})} 
\newcommand{\Sec}[1]{Sec.~\ref{#1}}  
\newcommand{\Fig}[1]{Fig.~\ref{#1}} 
\newcommand{\Tab}[1]{Table~\ref{#1}}
\newcommand{\tenx}[1]{\times 10^{#1}}

\begin{document}

\preprint{YITP-SB-29-13, FERMILAB-PUB-13-377-A-T, MCTP-13-27}

\title{Constraining Light Dark Matter with \\ Diffuse X-Ray and Gamma-Ray Observations}

\author{Rouven Essig}
\affiliation{C.N.~Yang Institute for Theoretical Physics,\\ Stony Brook University, Stony Brook, NY 11794, USA}

\author{ Eric Kuflik}
\affiliation{Raymond and Beverly Sackler School of Physics and Astronomy,\\ Tel-Aviv University, Tel-Aviv 69978, Israel}

\author{Samuel D. McDermott}
\affiliation{Michigan Center for Theoretical Physics,\\ University of Michigan, Ann Arbor, MI 48105, USA}
\affiliation{Theoretical Astrophysics Department\\Fermi National Accelerator Laboratory, P.O. Box 500, Batavia, IL 60510, USA}

\author{Tomer Volansky}
\affiliation{Raymond and Beverly Sackler School of Physics and Astronomy,\\ Tel-Aviv University, Tel-Aviv 69978, Israel}

\author{Kathryn M. Zurek}
\affiliation{Michigan Center for Theoretical Physics,\\ University of Michigan, Ann Arbor, MI 48105, USA}

\begin{abstract}
\noindent
We present constraints on decaying and annihilating dark matter (DM) in the 4~keV 
to 10~GeV mass range, using published results 
from the satellites HEAO-1, INTEGRAL, COMPTEL, EGRET, and the Fermi Gamma-ray Space Telescope.  
We derive analytic expressions for the gamma-ray spectra from various DM decay modes, and find lifetime constraints in the range $10^{24}-10^{28}$~sec, depending on the DM mass and decay mode.  We map these constraints onto the parameter space for a 
variety of models, including a hidden photino that is part of a kinetically mixed hidden sector, a gravitino with R-parity violating decays, a 
sterile neutrino, DM with a dipole moment, and a dark pion. 
The indirect constraints on sterile-neutrino and hidden-photino DM are found to be more powerful than other experimental or astrophysical probes in some parts of parameter space.
While our focus is on decaying DM, we also present constraints on DM annihilation to electron-positron pairs.  
We find that if the annihilation is $p$-wave suppressed,  the galactic diffuse constraints are, depending on the DM mass and velocity at recombination,  more powerful than the constraints from the Cosmic Microwave Background. 
\end{abstract}
\maketitle

\tableofcontents

\section{Introduction}
\label{sec:introduction}

A wide variety of precision astrophysical and cosmological observations have corroborated the existence of dark matter (DM),   
without providing any conclusive indications of its nature or its non-gravitational couplings to the Standard Model (SM).  
For the past 30 years, a broad experimental program has attempted to uncover the DM properties. However, the vast majority of the existing experiments search either for Weakly Interacting Massive Particles (WIMPs) or for axions, overlooking other theoretically viable and motivated possibilities.    One  interesting possibility is light dark matter (LDM) 
in the keV to 10~GeV mass range.    
In this paper, we focus on such DM and study constraints from existing indirect searches.   

A large class of models can accommodate DM with sub-GeV masses, see, {\it e.g.},~\cite{Boehm:2003bt,Boehm:2003ha,Pospelov:2007mp,Hooper:2008im,Feng:2008ya,Kaplan:2009ag,Essig:2010ye,Falkowski:2011xh,Essig:2011nj}.  
Such DM can be probed at 
colliders~\cite{Borodatchenkova:2005ct,Petriello:2008pu,Gershtein:2008bf,Goodman:2010ku,Fox:2011fx,bfactories}, at 
direct detection experiments~\cite{Essig:2011nj,Essig:2012yx,Graham:2012su}, 
and at proton- and electron-beam dumps~\cite{Batell:2009di,deNiverville:2011it,deNiverville:2012ij,Dharmapalan:2012xp,Izaguirre:2013uxa,Diamond:2013oda}.  
Constraints from the Cosmic Microwave Background (CMB) already limit the s-wave DM annihilation cross section to SM matter to be below that of a thermal WIMP, for DM masses below $\sim 7$~GeV~\cite{Galli:2009zc,Slatyer:2009yq,Finkbeiner:2011dx,Galli:2011rz}.   

While DM decays are less constrained by early Universe cosmology, stringent constraints can be placed on decaying DM from observations of the galactic and extra-galactic 
diffuse X-ray or gamma-ray background.  
The lifetime of Weak-scale DM is constrained from observations with the Fermi Large Area Telescope (Fermi LAT) to $\tau \gtrsim 10^{26} \mbox{ sec}$~\cite{Ackermann:2012qk,Ackermann:2012rg,Gomez-Vargas:2013bea,Papucci:2009gd,Cirelli:2009dv}, many orders of magnitude larger than the age of the Universe.  For DM below $\mathcal{O}(100\mbox{ MeV})$, the usual gamma ray constraints from the Fermi LAT do not apply, although the instruments on several other satellites (listed in Table~\ref{tab:jint} below) are sensitive to photons with energies well below a GeV.  The available data cover a photon energy range from 10's of GeV down to a few keV, providing the possibility of exploring a much broader range of DM candidates than WIMPs.  
Indeed, some of these data have already been utilized to constrain LDM, see, {\it e.g.},~\cite{Kribs:1996ac,Boyarsky:2008mx, Bertone:2007aw, Boyarsky:2006ag,Boyarsky:2006hr,Boyarsky:2007ge,Yuksel:2007dr,Cembranos:2008bw,GomezVargas:2011ph,Stecker:1988ar,Stecker:1978du,Pullen:2006sy}.  
Sterile neutrinos with a mass  $\sim \cO(1- 10 \kev)$ are a particularly popular candidate and their constraints have been explored in, {\it e.g.},~\cite{Boyarsky:2008mx,Boyarsky:2006ag,Boyarsky:2006hr,Boyarsky:2007ge,Masso:1999wj,Abazajian:2001vt,Boyarsky:2005us,Boyarsky:2006zi,RiemerSorensen:2006fh,Watson:2006qb,Abazajian:2006jc,RiemerSorensen:2006fh,Kusenko:2009up,silk,Yuksel:2007xh,Boyarsky:2006fg,Boyarsky:2006kc,Boyarsky:2009ix}.  
Below a few keV, thermal DM candidates become too warm to adequately explain the formation of structure in the Universe, so that such candidates necessarily have a mass above the lower energy bound accessible by these satellite experiments.

The goal of this paper is to derive constraints on light DM candidates in the keV to 10 GeV mass range, using the diffuse photon spectra data listed in Table~\ref{tab:jint}. 
We update and extend several results in the literature.  Taking a largely model-independent approach, we discuss a wide range of DM decay topologies.  We consider photons that are 
produced directly in the decay or from final state radiation (FSR) of charged particles that are 
produced in two- or three-body decays.   We map our results onto several known LDM models, and show limits on the corresponding  model parameter space.  For example, we consider constraints on a kinetically mixed supersymmetric hidden sector (with the hidden photino decaying to $\tilde G \gamma$ or $\tilde G e^+ e^-$, with $\tilde G$ the gravitino) and a sterile neutrino (with the sterile neutrino decaying to a neutrino and a photon).  While the constraints we derive are robust, they are based on published data.   Consequently, they can easily be improved by optimizing the search regions and taking better account of the signal and background fitting. 

While our focus is on decaying DM, we also consider annihilating DM.   
A thermal relic with a $p$-wave (or velocity suppressed) annihilation cross section is less constrained from CMB data 
than s-wave annihilation, 
since DM is cold at the CMB epoch.  For this case, we find that the limits from the diffuse 
background can be more constraining than the CMB. 

\begin{table}[t]
\begin{center}
{\small
\begin{tabular}{|c||c|c||c|c|c|c|c|c|c|} \hline
Experiment & $E_{\rm min}$ & $E_{\rm max}$ &  $\Omega$ & $J^{\rm NFW}_{D(A)}$ & $J^{\rm Moore}_{D(A)}$ & $J^{\rm IsoT}_{D(A)}$ & $J^{\rm Ein, 0.17}_{D(A)}$ & $J^{\rm Ein, 0.12}_{D(A)}$ & $J^{\rm Ein, 0.20}_{D(A)}$ \\ \hline
HEAO-1 \cite{Gruber:1999yr} &  $4 \kev $&$ 30\kev_{{}_{}}$  & \begin{minipage}{1.1in} \rule{0pt}{2.6ex} $58\leq\ell\leq109^\circ \cup 238\leq\ell\leq289^\circ $, $ 20^\circ\leq|b|\leq90^\circ_{{}_{}}$ \end{minipage}   & 
\begin{minipage}{.4in} \rule{0pt}{2.6ex} 3.88 (2.16) \end{minipage} &  
\begin{minipage}{.4in} \rule{0pt}{2.6ex} 4.06 (2.22) \end{minipage} &
\begin{minipage}{.4in} \rule{0pt}{2.6ex} 4.33 (2.24) \end{minipage} &  
\begin{minipage}{.4in} \rule{0pt}{2.6ex} 3.79 (2.09) \end{minipage} &  
\begin{minipage}{.4in} \rule{0pt}{2.6ex} 3.76 (2.05) \end{minipage} &  
\begin{minipage}{.4in} \rule{0pt}{2.6ex} 3.80 (2.11) \end{minipage} \\ \hline
INTEGRAL \cite{Bouchet:2008rp} & $20 \kev $ & $1\mev_{{}_{}}$ & \begin{minipage}{.9in} \rule{0pt}{2.6ex}  $|\ell|\leq30^\circ$, $|b|\leq15^\circ_{{}_{}}$ \end{minipage} & 
\begin{minipage}{.4in} \rule{0pt}{2.6ex} 3.65 (18.4) \end{minipage} &
\begin{minipage}{.4in} \rule{0pt}{2.6ex} 3.80 (24.4) \end{minipage} &
\begin{minipage}{.4in} \rule{0pt}{2.6ex} 2.77 (5.08) \end{minipage} &  
\begin{minipage}{.4in} \rule{0pt}{2.6ex} 4.20 (30.9) \end{minipage} &  
\begin{minipage}{.4in} \rule{0pt}{2.6ex} 4.73 (59.9) \end{minipage} &
\begin{minipage}{.4in} \rule{0pt}{2.6ex} 3.95 (23.2) \end{minipage} \\ \hline
COMPTEL \cite{COMPTEL} & $1 \mev $&$ 15\mev$ & \begin{minipage}{.9in} \rule{0pt}{2.6ex}  $|\ell|\leq60^\circ$, $|b|\leq20^\circ_{{}_{}}$ \end{minipage}   & 
\begin{minipage}{.4in} \rule{0pt}{2.6ex} 6.82 (23.1) \end{minipage} & 
\begin{minipage}{.4in} \rule{0pt}{2.6ex} 7.03 (29.1) \end{minipage} & 
\begin{minipage}{.4in} \rule{0pt}{2.6ex} 5.91 (8.69) \end{minipage} & 
\begin{minipage}{.4in} \rule{0pt}{2.6ex} 7.48 (36.4) \end{minipage} & 
\begin{minipage}{.4in} \rule{0pt}{2.6ex} 8.10 (66.0) \end{minipage} & 
\begin{minipage}{.4in} \rule{0pt}{2.6ex} 7.19 (28.3) \end{minipage} \\ \hline
EGRET \cite{Strong:2003ey} &$20 \mev $&$ 6\gev_{{}_{}}$  & \begin{minipage}{1.1in} \rule{0pt}{2.6ex} $0\leq\ell\leq360^\circ $, $ 20^\circ\leq|b|\leq60^\circ_{{}_{}}$ \end{minipage}  & 
\begin{minipage}{.4in} \rule{0pt}{2.6ex} 13.0 (10.9) \end{minipage} & 
\begin{minipage}{.4in} \rule{0pt}{2.6ex} 13.5 (11.0) \end{minipage} & 
\begin{minipage}{.4in} \rule{0pt}{2.6ex} 14.0 (10.1) \end{minipage} & 
\begin{minipage}{.4in} \rule{0pt}{2.6ex} 12.9 (11.5) \end{minipage} & 
\begin{minipage}{.4in} \rule{0pt}{2.6ex} 13.0 (12.0) \end{minipage} & 
\begin{minipage}{.4in} \rule{0pt}{2.6ex} 12.9 (11.3) \end{minipage}  \\ \hline
Fermi \cite{FermiLAT:2012aa} &  $200 \mev $&$ 10 \gev_{{}_{}}$  & \begin{minipage}{1.1in} \rule{0pt}{2.6ex} $0\leq\ell\leq360^\circ $, $ 8^\circ\leq|b|\leq90^\circ_{{}_{}}$ \end{minipage}   & 
\begin{minipage}{.4in} \rule{0pt}{2.6ex} 21.9 (22.0) \end{minipage} & 
\begin{minipage}{.4in} \rule{0pt}{2.6ex} 22.8 (22.5) \end{minipage} & 
\begin{minipage}{.4in} \rule{0pt}{2.6ex} 23.3 (17.9) \end{minipage} & 
\begin{minipage}{.4in} \rule{0pt}{2.6ex} 22.0 (25.4) \end{minipage} & 
\begin{minipage}{.4in} \rule{0pt}{2.6ex} 22.3 (28.5) \end{minipage} & 
\begin{minipage}{.4in} \rule{0pt}{2.6ex} 21.9 (24.0) \end{minipage} \\ \hline
\end{tabular}
}
\end{center}
\caption{\label{tab:jint} Energy ranges, solid angles, and values of $J_D~(J_A)$ for various  DM density profiles. The NFW profile is taken from~\cite{Navarro:1995iw,Navarro:1996gj}, the Moore profile from \cite{Kazantzidis:2003hb},  and the cored isothermal profile can be found in~\cite{IsoT}. The profiles ``Ein, $\alpha$" are Einasto profiles~\cite{Einasto} with slope parameter $\alpha$.  }
\end{table}

The outline of the paper is as follows. In Sec.~\ref{sec:Indirect}, we review both the expected signals resulting from DM decays and annihilations as well as  the relevant gamma-ray and X-ray observatories (HEAO-1, INTEGRAL, COMPTEL, EGRET, Fermi).   We further  discuss our methods for placing the limits on such DM.  In Sec.~\ref{sec:models}, we discuss models of decaying light DM such as decaying gravitinos, sterile neutrinos, and hidden photinos.  For each model we map the lifetime constraints onto constraints of the model parameter space.  In Sec.~\ref{sec:spectra}, we take a model-independent approach and constrain the lifetime for various decay topologies. 
Sec.~\ref{sec:annihilating} is devoted to constraints on the annihilation cross-section of light DM to electron-positron pairs.  We conclude in Sec.~\ref{sec:conclusions}.

\section{Constraining Light Dark Matter with Diffuse Photons}
\label{sec:Indirect}

In this section, we discuss the data and the statistical methodÊwe use to place constraints on decaying and annihilating LDM.  We begin  with a brief review of  the expected signal rate.

\subsection{Flux from Dark Matter Decays and Annihilations}

Given a DM annihilation or decay spectrum, $dN_\gamma/dE_\gamma$, and a galactic DM density profile, $\rho(r)$, the  galactic contribution to the differential photon flux per unit energy is given by,
 \beq \label{flux}
\frac{d \Phi_{\gamma,{\rm G}}}{ d E} = \frac{1}{2^{\alpha-1}}\frac{r_\odot }{4\pi} \frac{\rho_\odot}{m_{\rm DM}} \Gamma_{D,A} \frac{d N_\gamma}{d E} J_{D,A}^{}\,.
\eeq
Here $r_\odot\simeq 8.5$ kpc is the Sun's distance from the Galactic center, $\rho_\odot = 0.3 \mathrm{\ GeV/cm^3}$ is the local DM density, $\alpha = 1$ (2) for DM decays (annihilations), $\Gamma_D$ is the decay rate, $\Gamma_A = (\rho_\odot/m_{\rm DM})\langle \sigma v\rangle$ is the thermally averaged annihilation rate, and
\beq
J_{D,A}= \int_{\rm l.o.s.} \frac{ds}{r_\odot} \bL \frac{\rho(s)}{\rho_\odot} \bR^{\alpha} d\Omega,
\eeq
is a dimensionless quantity that describes the density of decays or annihilations along the line-of-sight (l.o.s.) and over the solid angle $\Omega$.  We will present results assuming $\rho(s)$ follows the NFW DM density profile~\cite{Navarro:1995iw,Navarro:1996gj}, but in Table~\ref{tab:jint} we also list values of $J_{D,A}$ for other DM density profiles for each experimental survey region.  Our results can thus be easily rescaled.  Note that the choice of $\rho(s)$ becomes less important for survey regions farther from the galactic plane
and also less important for decaying compared to annihilating DM.

In addition to the contribution to the photon flux from DM decays in the Milky Way halo, there is a contribution arising from the smooth distribution of DM throughout the whole Universe (see, {\it e.g.},~\cite{Masso:1999wj,Abazajian:2001vt,Boyarsky:2005us,Bertone:2007aw}).  A photon produced at redshift $z$ that is detected with energy $E$ was emitted with energy $E(z) = E (1+z)$.  Such a photon was emitted at a comoving distance, $\chi(z)$, with
\beq
\frac{d\chi(z)}{dz} = \frac{1}{(1+z)^{3/2}}\,\frac{1}{a_0H_0\sqrt{\Omega_m(1+\kappa (1+z)^{-3})}}\,, 
\eeq
where $\kappa=\Omega_\Lambda/\Omega_m\sim 3$ and a flat Universe, $\Omega_m + \Omega_\Lambda=1$, is assumed.  The extragalactic photon spectrum arising from DM decays at redshift $z$ is given by $dN/dE(z)$, so that the measured flux is
\beq\label{eq:redshifted}
\frac{d^2\Phi_{\gamma, EG}}{d\Omega dE} = \frac{1}{4\pi}\,\frac{\Gamma \Omega_{\rm DM} \rho_c}{m_{\rm DM} a_0H_0 \sqrt{\Omega_m}}\;
\int_0^\infty\,dz\, \frac{dN}{dE(z)}\, \frac{1}{(1+z)^{3/2}}\, \frac{1}{\sqrt{1+\kappa (1+z)^{-3}}}\,.  
\eeq
Because the photon flux from DM decays scales linearly with the DM density, this contribution is not very model dependent.
For $dN_\gamma/dE(z) = \delta(E(z) - m_{\rm DM}/2)$, this reduces to the case that is usually considered, namely DM decaying to a redshifted monochromatic gamma-ray line,
\beq
\frac{d^2\Phi_{\gamma, EG}}{d\Omega dE} = \frac{1}{4\pi}\,\frac{\Gamma \Omega_{\rm DM} \rho_c}{m_{\rm DM} H_0 \sqrt{\Omega_m}}\, 
\left(\frac{2}{m_{\rm DM}}\right)\, \sqrt{\frac{2 E}{m_{\rm DM}}} \, \frac{1}{\sqrt{1+\kappa (2E/m_{\rm DM})^3}}\,.
\eeq
This effect implies that  the spectral shape of a photon ``line'' from DM decays is  smeared to receive contributions from a continuous range of energies.

In principle, similar extragalactic contributions exist for the annihilating DM case.   However, the smooth part of extragalactic DM annihilation is  subdominant compared to the galactic contribution and may be safely ignored.  On the other hand,  extragalactic annihilations resulting from DM substructure at low redshift may contribute a significant amount to the photon flux since it scales as the square of the DM density~\cite{Ullio:2002pj}.  Since this contribution is 
not well known~\cite{Taylor:2002zd}, we conservatively omit it from our analysis below.  

For DM decay or annihilation to final states that include electrons or positrons, there are other potentially important contributions to the diffuse photon flux.  The electrons and positrons can inverse Compton scatter (ICS) starlight, infrared, or CMB photons, or produce synchrotron radiation.   The precise contribution to the diffuse flux, however, is model dependent and requires detailed knowledge of  the galactic and extragalactic magnetic fields as well as the diffusion properties of the electrons in our Galaxy.   In order to present conservative bounds and to avoid 
significant systematic uncertainties, we do not include these contributions. 

When stable charged particles (like electrons) appear as decay or annihilation products, photons will be emitted through final state radiation (FSR).  We use the Altarelli-Parisi splitting function, 
\beq \label{apsf}
\frac{d \Gamma_{\rm FSR}}{d E_\gamma}=\frac{\alpha_{\rm EM} \Gamma_{D,A} }{2\pi} \ln (Q/m_f^2) \int  \frac{1+(1-E_\gamma/E_f)^2}{E_\gamma} \frac{d N}{d E_f} dE_f,
\eeq
to  estimate the photon spectrum, where $Q$ is the square of the momentum imparted to the photon, $\alpha_{\rm EM} \simeq 1/137$, and $d N/dE_f$  is the differential rate of decay or annihilation to the final state particle $f$.   For multiple charged-particles in the final state, we sum over the contributions.

\subsection{Data}

\begin{figure}[t!]
\begin{center}
\includegraphics[width=.9\textwidth]{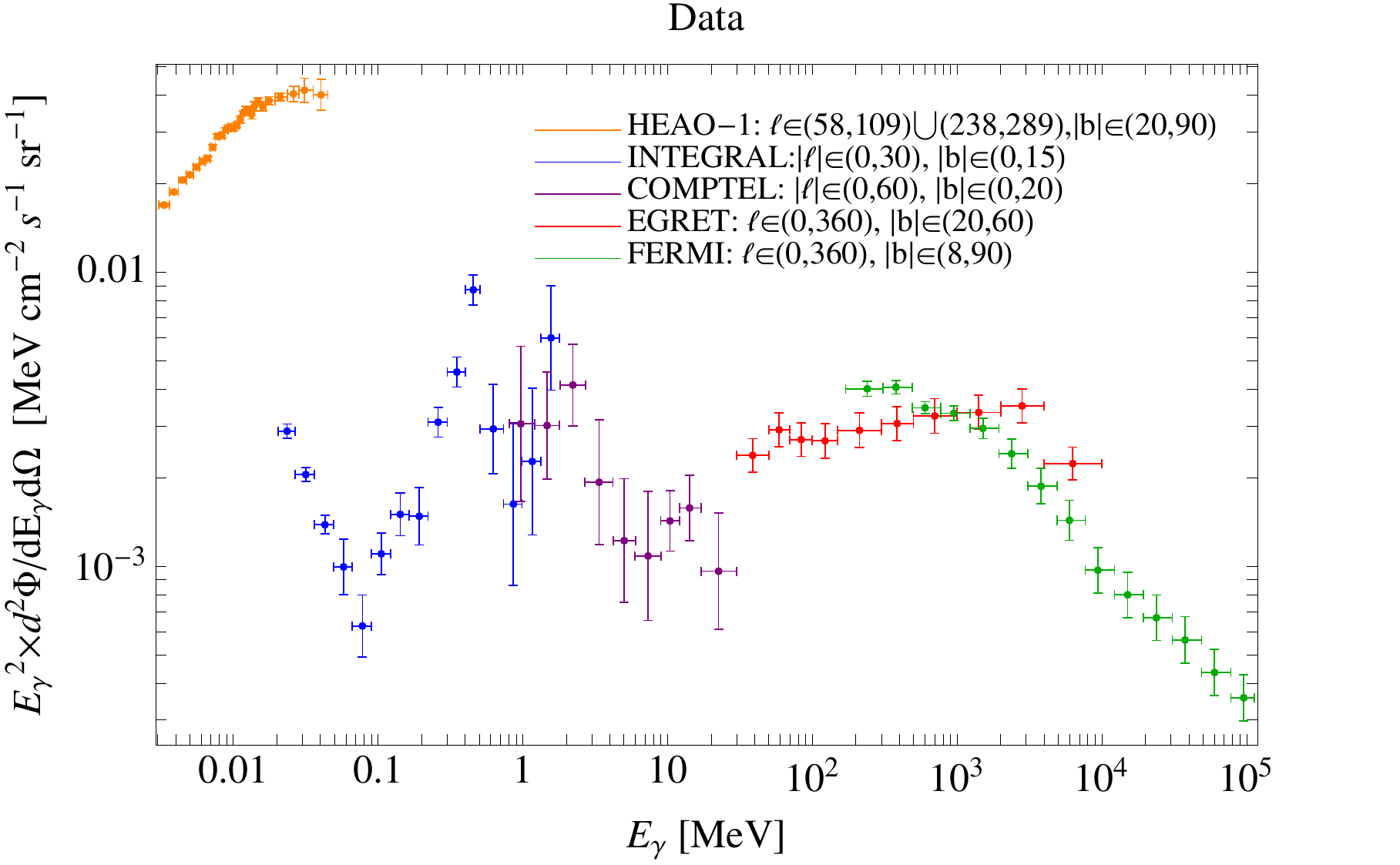}
\caption{The collected normalized dataset of photon fluxes used to place constraints on decaying and annihilating DM in this paper.   Data from HEAO-1~\cite{Gruber:1999yr} (orange), INTEGRAL~\cite{Bouchet:2008rp} (green), COMPTEL~\cite{COMPTEL} (blue),  EGRET~\cite{Strong:2003ey} (red), and Fermi~\cite{FermiLAT:2012aa} (yellow) are shown.  All error bars are statistical, except for the EGRET and Fermi datasets, where the dominant systematic uncertainties are shown. We omit the INTEGRAL 511 keV line both in this figure and in our analysis. Note that the various datasets span different regions of the sky and should therefore not be compared with each other; they appear together on this plot only for convenience.  }
\label{fig:data}
\end{center}
\end{figure}

We place constraints on LDM using the data summarized in \Tab{tab:jint} and shown in \Fig{fig:data}.  We emphasize that 
\emph{none of the datasets have been optimized for LDM searches}.  
It is therefore likely that significantly stronger constraints may be achieved with dedicated analyses.

As mentioned above, we assume an NFW profile in all cases, but the results can easily be rescaled for other profiles using the information in \Tab{tab:jint}. For the inner-galaxy data from INTEGRAL or COMPTEL, the bounds from decaying DM can be adjusted by up to $\cO(30\% )$; using high-latitude data, the difference is typically less than $\cO(10\%)$.  In contrast, the expected photon flux from DM annihilations near the galactic center can change by up to an order of magnitude for different choices of the density profile.

For our analysis we use the following datasets:
\begin{itemize}
\item{\bf HEAO-1}.
We use data from observations of 3--50 keV photons made with the A2 High-Energy Detector on HEAO-1~\cite{Gruber:1999yr}.  Other datasets from the experiment are significantly weaker than those from the INTEGRAL experiment discussed below.   To avoid point source contamination, the observations come from regions of the sky $20^\circ$ above the galactic plane.  
As is clear from \Tab{tab:jint}, the constraints from this sky region are not very sensitive to the DM density profile. 

\item{\bf INTEGRAL}.
We use data from observations of 20 keV to 2 MeV photons from the region $|\ell|<30^\circ$ and $|b|<15^\circ$ obtained with the SPI instrument onboard INTEGRAL~\cite{Bouchet:2008rp}. The quantity $J$ changes by up to $ \cO(30\%)$ in the decaying case, for different choices of density profile. The excellent energy resolution allows us to remove the well-resolved 511 keV line in our analysis. 

\item{\bf COMPTEL}.
We use the COMPTEL data from \cite{COMPTEL}. These observations are obtained by averaging over the sky at latitudes  $|\ell|\leq60^\circ$ and  $|b|<20^\circ$.   Compared to the INTEGRAL region of interest, the model predictions are about half as sensitive to the density profile at these galactic latitudes. We find an ${\cal O}(20\%)$ uncertainty for DM decay bounds due to the DM density profile.

\item{\bf EGRET}. 
We use the data shown in panel E of Fig.~2 in~\cite{Strong:2003ey}, which lies in the 20~MeV to 10~GeV range at intermediate latitudes,  
$0\leq\ell\leq360^\circ,20^\circ\leq|b|\leq60^\circ$. 
Our results are sensitive only at the few-percent level to the DM density profile.

\item{\bf Fermi}. 
We use data from the upper panel of Fig.~12 of \cite{FermiLAT:2012aa}, with $0<\ell<360^\circ$ and $8^\circ<|b|<90^\circ$, between 200~MeV--10~GeV. We choose these latitudes to enhance the signal to background ratio while minimizing the uncertainty in the DM profile. The resulting decay bounds are only $ \cO(5\%)$ sensitive to varying the DM density profile.
\end{itemize}

\subsection{Statistical Methodology}\label{sec:stat}

Our goal is to obtain robust, conservative bounds using the above data sets.  We do this by requiring that the predicted count from the DM signal in each bin does not exceed the observed central value plus twice the error bar. In all cases we use the statistical uncertainties, except for EGRET and Fermi where we take the dominant systematic uncertainties.  These bounds could be significantly strengthened with dedicated searches in the future and by including fits to different astrophysical background components, {\it e.g.}, from astrophysical ICS.  In Appendix~\ref{sec:gof}, we show the improvement that could be obtained with a goodness-of-fit test that assumes knowledge of the various backgrounds.  The expected improvement varies between a factor of a few to an order of magnitude, but involves larger systematic uncertainties as the backgrounds are not precisely known.   For this reason, the results we present use only this simple test described above.


\section{Models of Decaying Light Dark Matter}
\label{sec:models}

In this section, we outline several simple scenarios that can accommodate LDM, and we place constraints on the model parameter space.   
The models below should be viewed as benchmarks that are not, however, complete.  In particular, we do not discuss the production mechanism that results in the observed relic abundance.  
In the next section, we will derive ``model-independent'' constraints, where the results are presented as generic constraints on the lifetime versus mass for a given decay topology.  

\subsection{Hidden Photino}
\label{sec:hidgaugino}

Consider a supersymmetric hidden sector, with an additional $\Udark$ gauge group~\cite{Hooper:2008im,ArkaniHamed:2008qn, Cheung:2009qd,Morrissey:2009ur,Ruderman:2009tj,Essig:2010ye}.   We assume that the SM and hidden sector can interact with each other through gauge kinetic mixing~\cite{Holdom:1985ag,Galison:1983pa}, 
\begin{eqnarray}
  \label{eq:kineticmix}
  -\frac{\epsilon}{2}\int d^2\theta\  \Wdark \W_Y\,,
\end{eqnarray}
where $\Wdark$ ($\W_Y$) are the supersymmetrized field strength of the hidden gauge group (hypercharge).  
The value of $\epsilon$ may naturally be of order $10^{-3} - 10^{-4}$ when generated by integrating out heavy fields charged under both sectors.  Conversely, if Eq.~\eqref{eq:kineticmix}  results from  higher dimensional operators, $\epsilon$ can be significantly smaller, as we will assume below in order to obtain MeV-GeV masses.   

\begin{figure}[t]
\begin{center}
\includegraphics[width=0.47\textwidth]{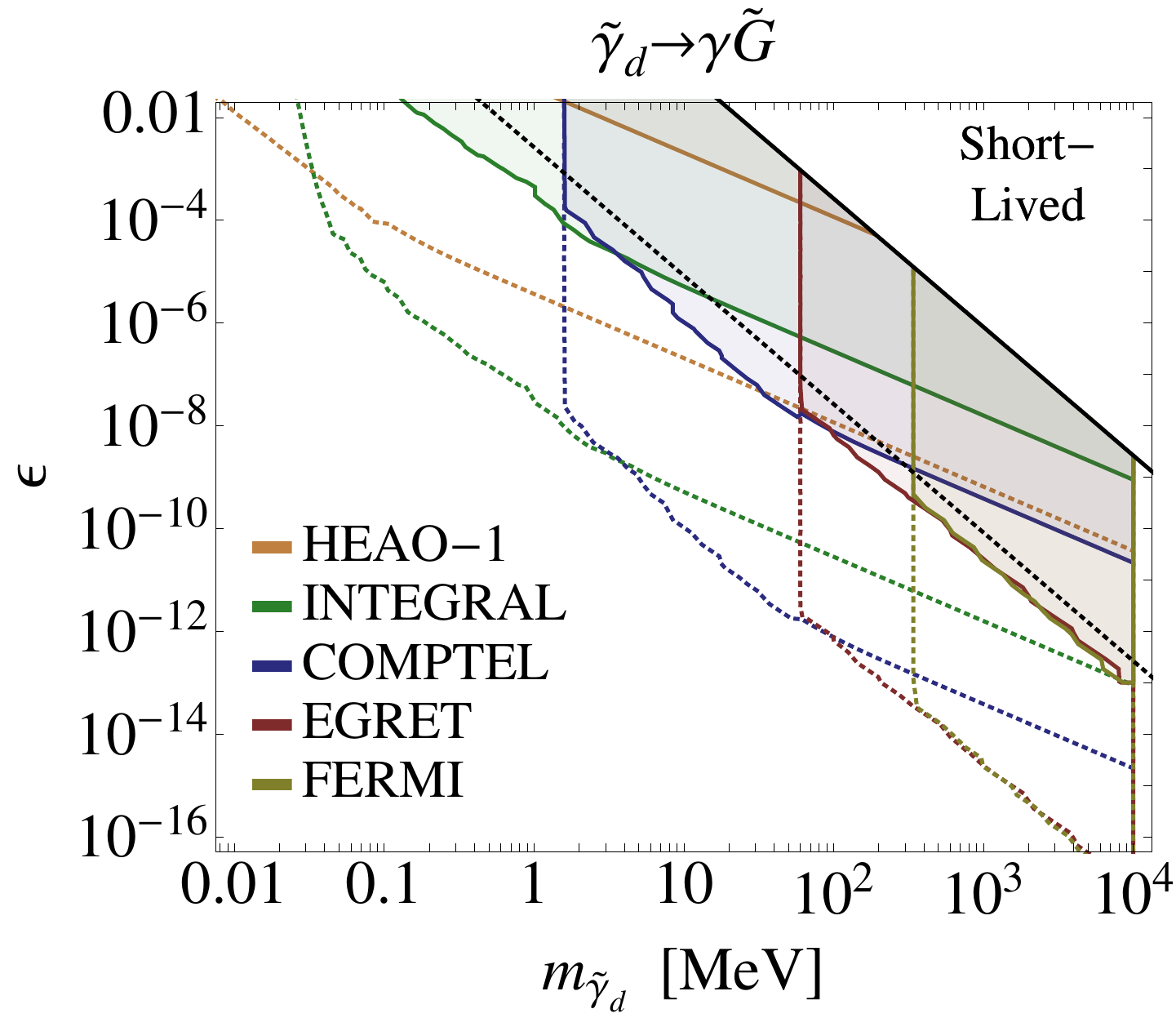}~~~~~ \includegraphics[width=0.45\textwidth]{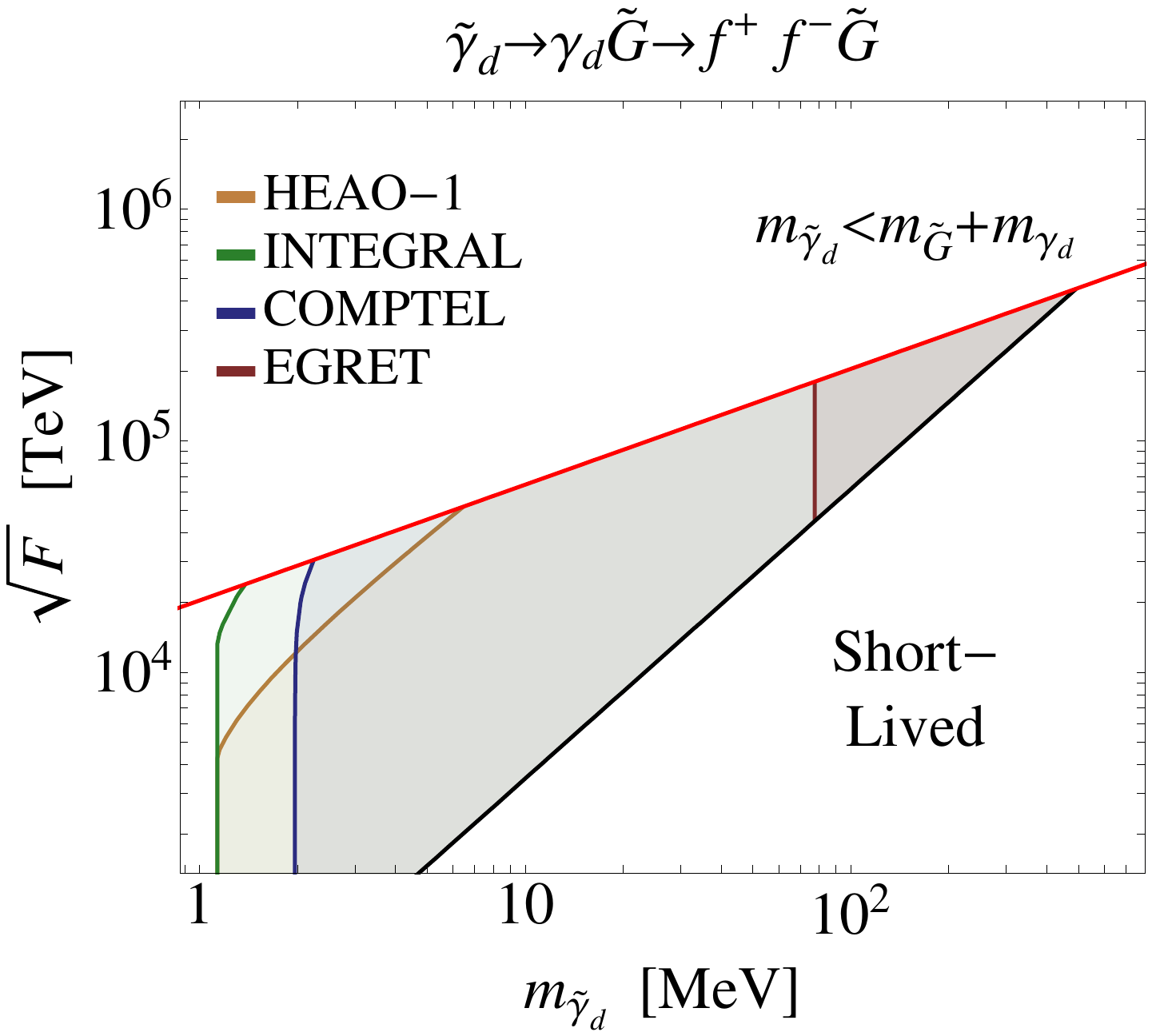}
\caption{Constraints on hidden  photino decay to {\bf left:} gravitino and  photon and {\bf right:} gravitino and hidden photon (with the latter taken to have mass $m_{\gamma_d} = 0.9 m_{\wt \gamma_d}$ and going to final state $f^+ f^-$, with $f=e,\mu$ or $\pi$). In the left plot, the solid (dotted) lines are with $\sqrt{F}=10^4~(10^2)\tev$.  
The constraints are derived from the diffuse gamma- and X-ray data taken from HEAO-1 (orange), INTEGRAL (green), COMPTEL (blue), EGRET (red), and Fermi (yellow). In the ``Short-Lived'' region the DM lifetime is shorter than the age of Universe.  Above the solid 
red line, the hidden photino is stable.}
\label{fig:darkphoton}
\end{center}
\end{figure}

\begin{figure}[t]
\begin{center}
\includegraphics[width=0.49 \textwidth]{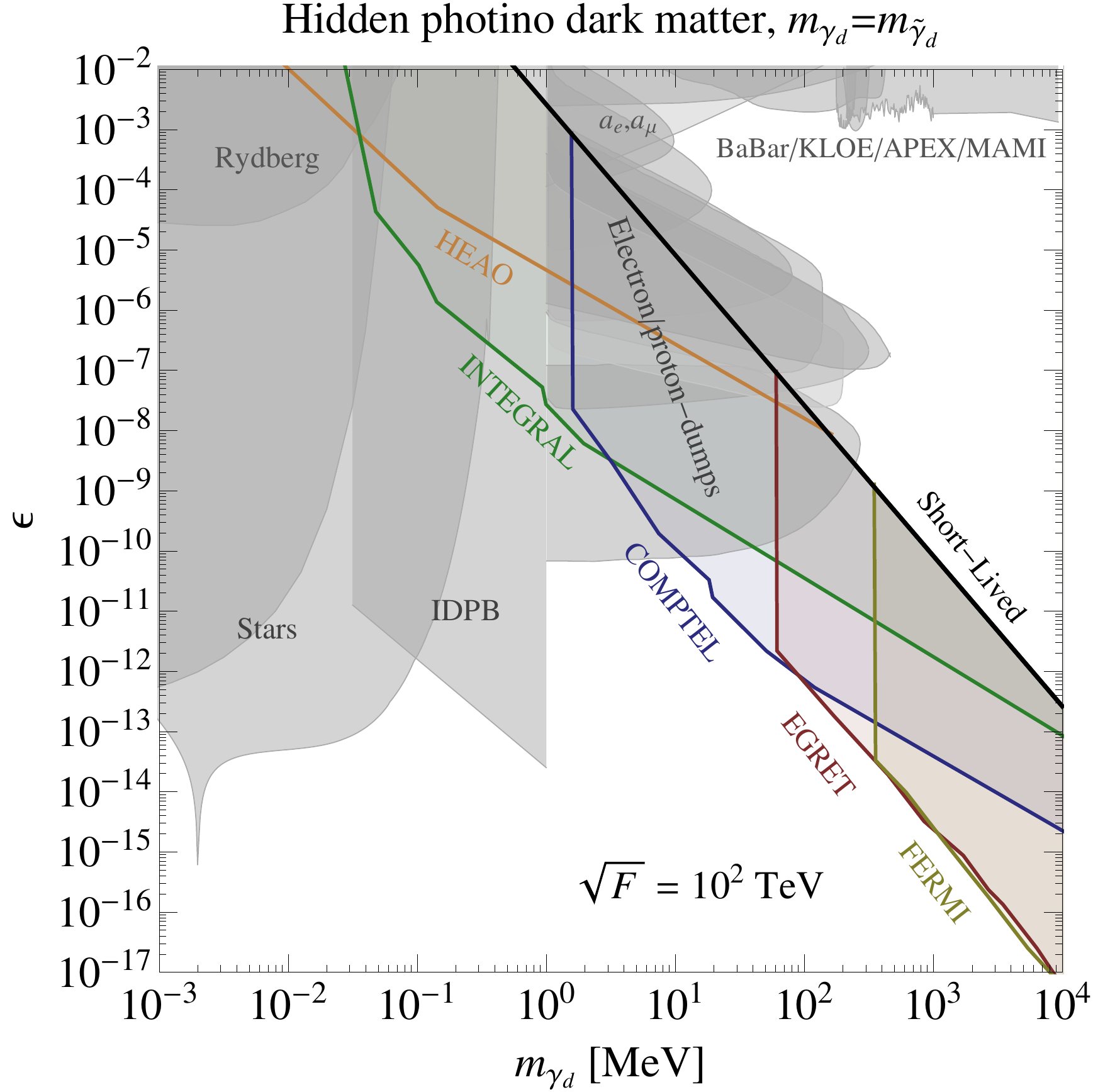}~~~
\includegraphics[width=0.49 \textwidth]{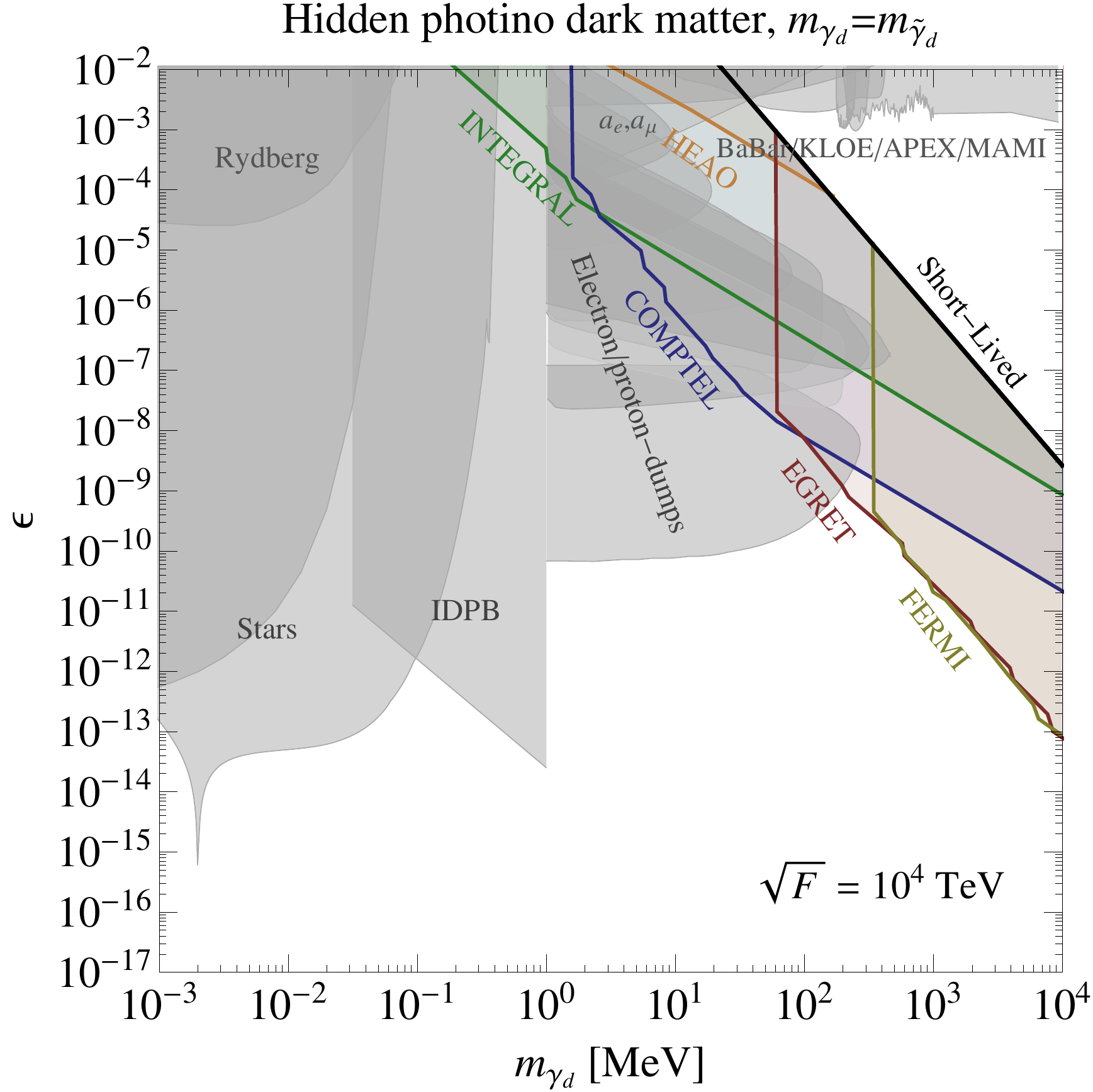}
\caption{Constraints on a hidden  photon in the hidden photino DM model for the case where the hidden photino decays to a photon and a gravitino, $\tilde \gamma_d \to \gamma \tilde G$, and with $\sqrt{F}=100\tev$ ({\bf left}) 
or $\sqrt{F}=10^4\tev$ ({\bf right}).  Gray shaded regions indicate constraints from beam-dump, fixed-target, and colliding beam experiments, stars, precision measurements, and from the intergalactic diffuse photon background (IDPB), 
while the colored regions show the gamma- or X-ray constraints as in Fig.~\ref{fig:darkphoton}. 
In the ``Short-Lived'' region the DM lifetime is shorter than the age of Universe.
See text for more details.}
\label{fig:darkphoton2}
\end{center}
\end{figure}

An interesting possibility is to have the hidden gaugino play the role of DM.  To realize this, supersymmetry must be broken and communicated both to the visible and hidden sector.   If the communication occurs through gauge mediation, the breaking in the hidden sector may be significantly smaller than in the visible sector as supersymmetry breaking is transmitted to the hidden sector  through D-term mixing~\cite{Cheung:2009qd}.  As a consequence, the hidden photon mass is given by,
\begin{equation}
\label{eq:mdark}
\mdark^2 =  \epsilon\ \gdark \left< D_Y \right> \simeq \left (5 \unit{MeV} \right)^2 \left( \frac{\epsilon}{10^{-8}} \right)  \left( \frac{\gdark}{0.2} \right) \left( \frac{\sqrt{\left< D_Y \right>}}{50 \unit{GeV}} \right)^2\,,
\end{equation}
where $\langle D_Y \rangle = |\frac{g_Y v^2 c_{2 \beta}}{4}|$, $v = 246 \mbox{ GeV}$, and $\tan \beta = v_u/v_d$. 
  In such a case, $\gamma_d$ and $\tilde\gamma_d$ are nearly degenerate, and $\tilde\gamma_d$ can decay to the gravitino and either a photon or a hidden photon, depending on whether the latter is heavier or lighter than $\tilde \gamma_d$~\cite{Morrissey:2009ur,Ruderman:2009ta,Ruderman:2009tj,Essig:2010ye,Cohen:2010kn}.  
The hidden photino lifetime is,
\begin{equation}
\label{eq:decaytophoton}
\tau_{\dgaugino \rightarrow \gamma \tilde G} \simeq  \epsilon^{-2} \left(\frac{\mgaugino^5}{16 \pi F^2} \right)^{-1} \simeq 3\times 10^{23} \unit{sec}
\left( \frac{10^{-8}} {\epsilon} \right)^2 \left(\frac{10\unit{MeV}}{m_{\dgaugino}} \right)^5 \left(\frac{\sqrt{F}}{100\unit{TeV}} \right)^4\,,
\end{equation}
for the decay to the photon and gravitino.   This lifetime depends on several parameters, and can be much longer  for lighter DM if the exact relation, Eq.~\eqref{eq:mdark}, holds.   Of course, $\mgaugino$ can be controlled by some other dynamics and hence be independent of $\epsilon$.  
Similarly for  $\dgaugino \rightarrow \dgauge \, \tilde G$ we have,
\begin{eqnarray}
\label{eq:decaytodarkphoton}
\tau_{\dgaugino \rightarrow \dgauge \tilde G} &\simeq& \left(\frac{m_{\dgaugino}^5}{16 \pi F^2} \right)^{-1} \left(1- \nu_{\gamma_d}^2 \right)^{-4} \nonumber \\
&=& 3\times10^{20} \unit{sec}
\left(\frac{1\unit{MeV}}{m_{\dgaugino}} \right)^5 \left(\frac{\sqrt{F}}{10^4\unit{TeV}} \right)^4 \left(1-\frac{m_{\gamma_d}^2}{m_{\rm DM}^2} \right)^{-4}\,.
\end{eqnarray}
Here, a long lifetime requires a slightly larger SUSY breaking scale.
Note that the two possibilities lead to distinct indirect detection signals.   In the first case one expects a spectral line, while in the second the spectrum is dominated by the FSR photons from the kinematically accessible charged particles that arise from the decay of the hidden photon.

The constraints  for both cases are shown in Fig.~\ref{fig:darkphoton}.   In the case of the line, we show the bounds in the $\epsilon-m_{\widetilde \gamma_d}$ plane, taking two choices for $\sqrt{F}$. For the case where the photino decays via a hidden photon,  the constraints are presented on the $\sqrt{F}-m_{\widetilde \gamma_d}$ plane with the assumption $m_{\gamma_d}=0.9 m_{ \widetilde \gamma_d}$.  Above the solid red line, the hidden photino is stable.  The photon spectrum for a variety of different decay channels may be derived from \cite{Mardon:2009rc}. In both panels, the ``Short-Lived'' region indicates that the DM lifetime is shorter than the age of Universe.

   Assuming $m_{\gamma_d} \simeq m_{\dgaugino}$, additional constraints exist from beam-dump~\cite{Bjorken:2009mm,Andreas:2012mt}, fixed-target~\cite{Abrahamyan:2011gv,Merkel:2011ze}, and colliding-beam experiments~\cite{Aubert:2009cp}; precision measurements~\cite{Pospelov:2008zw}; stars~\cite{An:2013yfc,Redondo:2013lna}; and from the intergalactic diffuse photon background (IDPB). This final constraint is valid for hidden photons below $2 m_e \simeq$~1~MeV, as these can decay to three photons and contribute to the diffuse photon background~\cite{Redondo:2008ec,Yuksel:2007dr}.  For a summary of results see, {\it e.g.},~\cite{Hewett:2012ns}.  
These additional constraints are shown in Fig.~\ref{fig:darkphoton2} together with the limits derived here (and shown in the left panel of Fig.~\ref{fig:darkphoton}),  for the case where the hidden photon decays directly to a photon and a gravitino, $\tilde \gamma_d\to\gamma \tilde G$.  We note that some of these additional constraints are model dependent and may be evaded.

\subsection{Sterile Neutrino}
\label{sec:sterileneutrino}

\begin{figure}[t]
\begin{center}
\includegraphics{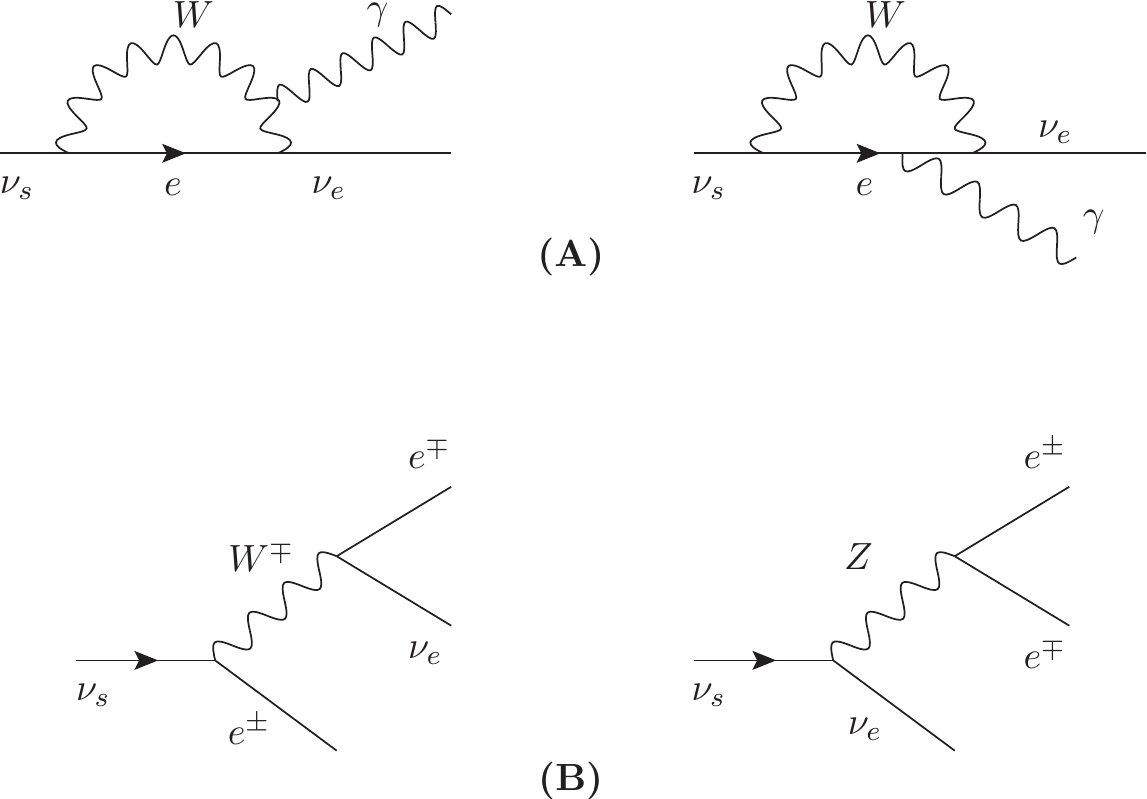}
\caption{ Decay channels for a sterile neutrino, $\nu_s$, through {\bf (A)} a two-body radiative process ($\nu_s \to \nu_\alpha \gamma$) and {\bf (B)} charge-  and neutral-current contributions to a three-body final state.}
\label{fig:sterile}
\end{center}
\end{figure}

\begin{figure}[htb]
\begin{center}
\includegraphics[width=0.49\textwidth]{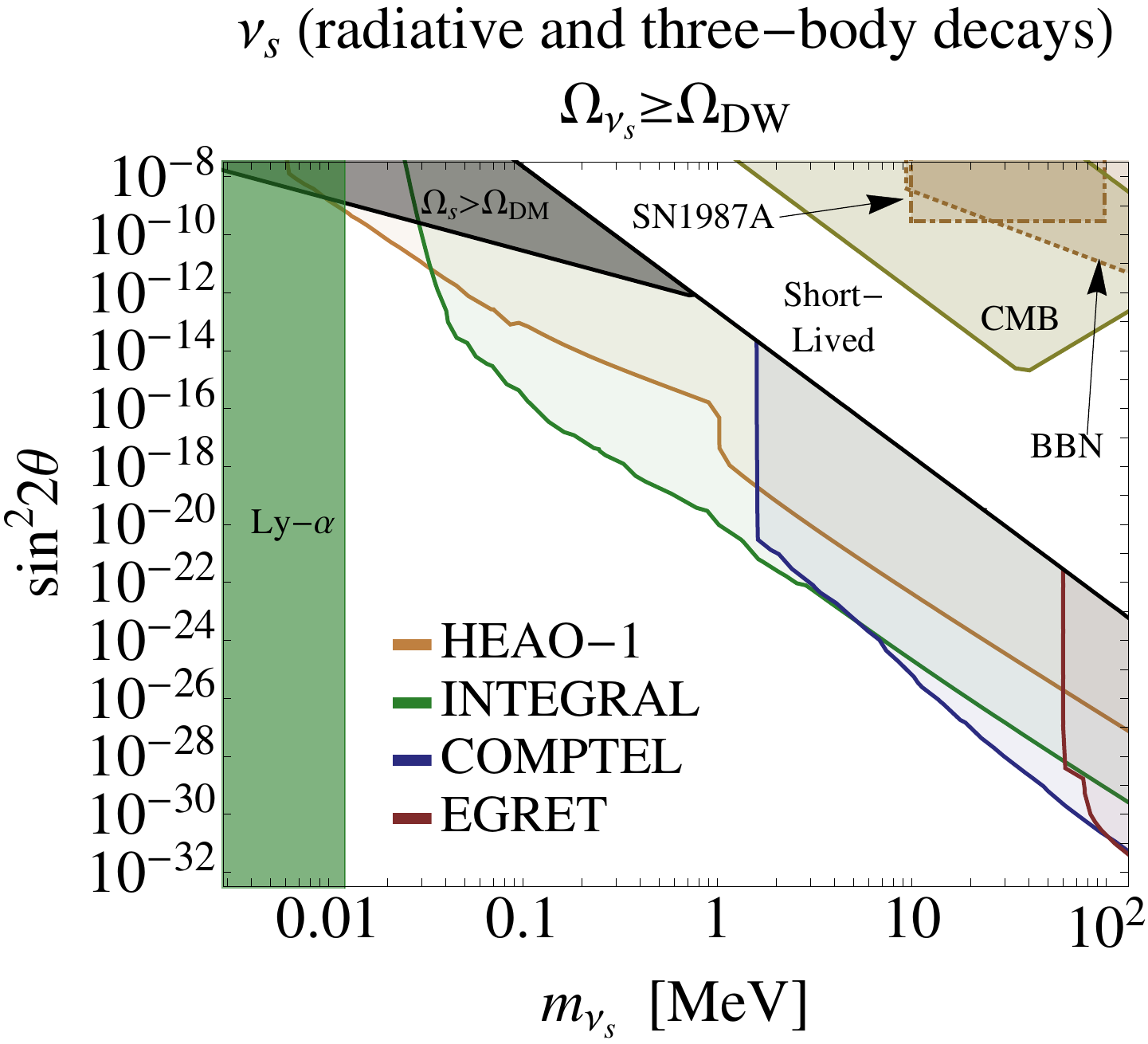}~~~~
\includegraphics[width=0.49\textwidth]{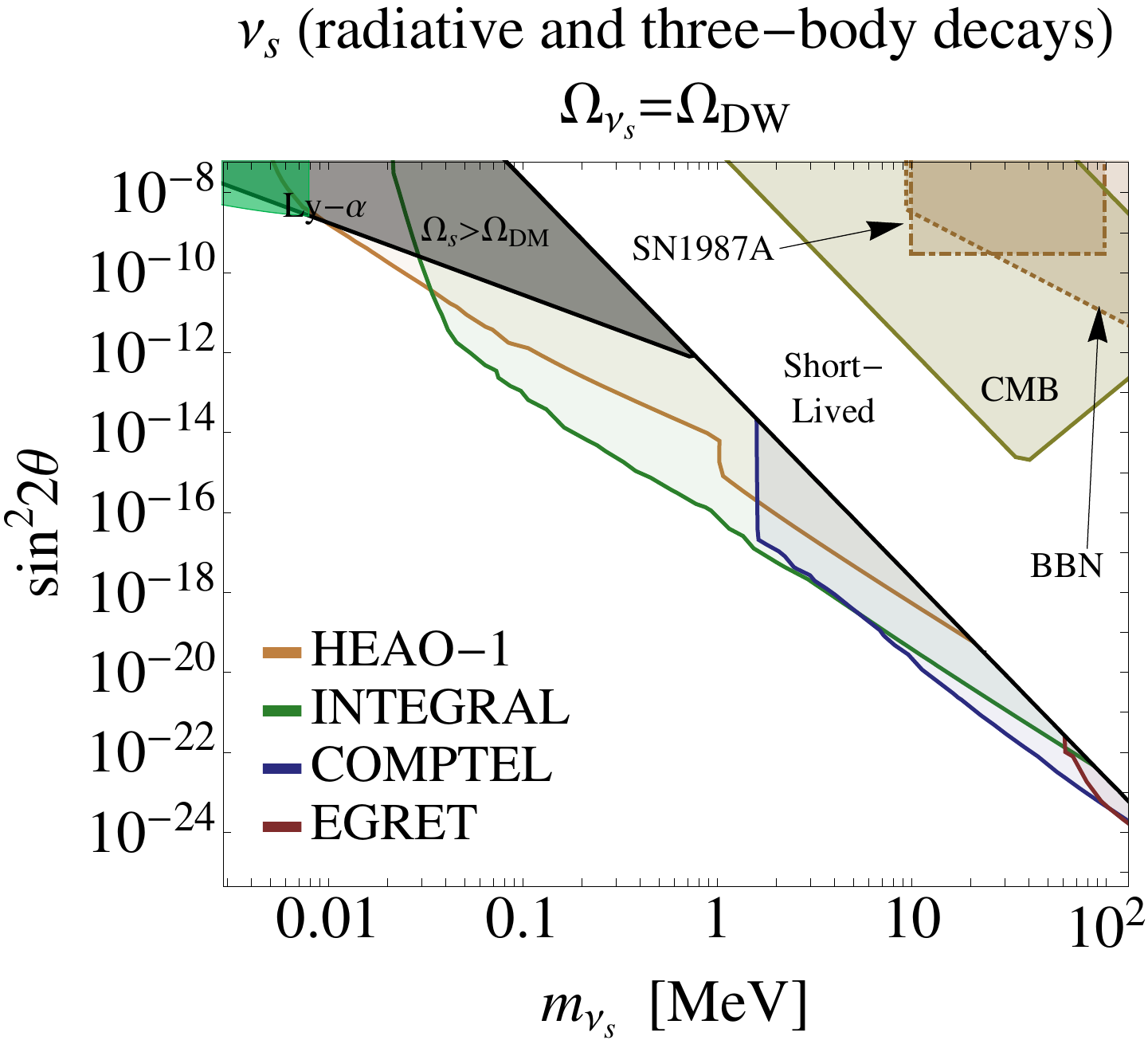}
\caption{Constraints on the sum of sterile-neutrino decay to $\gamma \nu$ and $\nu e^+ e^-$ using the decay widths in \Eqs{DMnugam} {nusterile3body}.  The constraints from the diffuse gamma- and X-ray data are HEAO-1 (orange), INTEGRAL (green), COMPTEL (blue), and EGRET (red). 
Within the solid black region, the neutrino energy density must be greater than the observed DM density. Above (below) the black solid line, the neutrino lifetime is shorter (longer) than the age of the Universe. Within the green boundaries, the sterile neutrino is ruled out by Ly-$\alpha$ forest data~\cite{Kusenko:2009up,silk}. Two cases for the sterile-neutrino energy density are assumed. In the {\bf left} plot, the density is assumed to precisely equal the DM energy density everywhere below the dark and light gray regions.  
In the {\bf right} plot, the density is determined by the (irreducible)  DW mechanism.}
\label{fig:sterileneutrino}
\end{center}
\end{figure}

Under certain circumstances, a sterile neutrino, $\nu_s$, may act as DM (for reviews, see~\cite{Kusenko:2009up,Boyarsky:2009ix}).   Due to its mixing with the active neutrinos, it may decay either via a 2- or  3-body channel.   The leading diagrams that contribute to these decay channels are shown in Fig.~\ref{fig:sterile}. In its simplest form, the theory at low energy is described by two parameters:
\begin{itemize}
\item $m_s$ - the sterile neutrino mass
\item $\sin\theta_\alpha$ - the mixing angle between $\nu_s$ and active neutrinos of flavor $\alpha$; in what follows, we will only consider $\nu_s - \nu_e$ mixing.
\end{itemize}
The mixing above can be induced, for example, in supersymmetric theories with a superpotential, $W = XLLE^c$. 
The two-body decay rate for a Majorana neutrino is given by~\cite{Shrock:1982sc}
\begin{equation}
\tau_{\nu_s \to \nu \gamma} \simeq \pL \frac{9\alpha_{\rm EM} \sin^2 \theta}{1024 \pi^4} G_F^2 m_\chi^5 \pR^{-1}  \simeq  1.8 \tenx{17} \s  \pL \frac{10\mev}{m_\chi} \pR^5 \pL  \frac{\sin \theta}{10^{-8}} \pR^{-2}\,,
\label{DMnugam}
\end{equation}
while the three-body decay rate is~\cite{Ruchayskiy:2011aa}
\begin{equation}
\label{nusterile3body}
\tau_{\nu_s\to \nu_\alpha e^+e^-} \simeq  \pL \frac{c_\alpha \sin^2 \theta}{96 \pi^3} G_F^2 m_\chi^5 \pR^{-1}  \simeq 2.4 \tenx{15} \s \pL \frac{10\mev}{m_\chi} \pR^5 \pL  \frac{\sin \theta}{10^{-8}} \pR^{-2}\,.
\end{equation}
Here the neutrino flavor $\alpha = e$, $c_\alpha= \frac{1+4 \sin^2 \theta_W + 8 \sin^4 \theta_W}4 \simeq 0.59$~\cite{Ruchayskiy:2011aa}, and we are only considering decays to $e^+e^-$ pairs. The resulting gamma-ray fluxes from both channels contribute at roughly similar levels once the splitting function is introduced. 

The relic abundance of sterile neutrinos is model dependent and varies according to the specific production mechanism  and dynamics in the early Universe.   An irreducible and UV-insensitive contribution to the abundance of sterile neutrinos arises from the so-called Dodelson-Widrow (DW) mechanism~\cite{Dodelson:1993je} in which the neutrinos are produced via oscillations.  Thus, in the absence of new dynamics at low temperature, one finds~\cite{Kusenko:2009up} 
\begin{equation}
\label{eq:omegas}
\Omega_s \gtrsim 0.25 \left(\frac{\sin^22\theta}{4.3\times 10^{-13}}\right)\left(\frac{m_s}{\textrm{MeV}}\right)^{1.8}\,.
\end{equation}
Additional contributions may arise from, {\it e.g.},~non-thermal production~\cite{Falkowski:2011xh} or due to an extended Higgs sector~\cite{Kusenko:2006rh,Petraki:2007gq}.

In order to place model-independent bounds on the parameter space of sterile neutrinos, we consider two different possibilities for the size of the sterile-neutrino relic abundance. First, we consider an unspecified UV mechanism that contributes to the DM density in those regions where the DM is under-abundant, setting $\Omega_{\nu_s} = \Omega_{\rm DM}$. Next, we assume the relic abundance is determined solely by the DW mechanism and, depending on the mixing angle and mass, $\Omega_{\nu_s}$ can be greater than or less than $ \Omega_{\rm DM}$.  
We show our bounds for both these cases in the left and right panel of Fig.~\ref{fig:sterileneutrino}, respectively, in the $m_{\nu_s}-\sin^22\theta$ plane.  In addition, we show existing bounds from the observation of the Lyman-$\alpha$ forest~\cite{silk} and the overclosure  region, in which  the neutrino density produced by the DW mechanism exceeds the observed DM density.   
We also show the region where the sterile-neutrino lifetime is shorter than the age of the Universe, and hence it cannot act as DM.   
Several additional constraints exist on sterile neutrinos, for example,  from the power spectrum of large scale structure~\cite{Smirnov:2006bu} and of the CMB~\cite{Smirnov:2006bu},  from BBN~\cite{Dolgov:2003sg}, and from Supernova-1987A~\cite{Kainulainen:1990bn}.  However, these constraints lie in the region where either the lifetime is too short or where the DM density is too high. 

\subsection{Gravitino Dark Matter}
\label{sec:gravitino}

Another interesting possibility is gravitino DM~\cite{Moroi:1993mb,Takayama:2000uz, Moreau:2001sr,Buchmuller:2007ui, Moroi:1995fs,Cheung:2011nn, Hall:2009bx, Hall:2013uga}.   The gravitino may be unstable on cosmological timescales and here we consider gravitino decays induced by R-parity violating (RPV) interactions~\cite{Takayama:2000uz,Moreau:2001sr, Buchmuller:2007ui}.  Since we are interested in light DM, we will focus on the RPV operator that allows the gravitino to decay to leptons, $W = \lambda_{ijk} \ell_i \ell_j e^c_k$.  A small coefficient $\lambda$ in the RPV vertex can ensure that the gravitino lifetime is  longer than the age of the Universe.

Gravitinos are typically produced in three processes~\cite{Moroi:1993mb}:  (i) gaugino scattering, dominantly at the re-heat temperature, (ii) freeze-out and decay of the lightest ordinary supersymmetric particle (LOSP, such as a neutralino), and (iii) freeze-in production from decays of visible sector particles, dominated at temperatures of order the superpartner masses.

\begin{figure}[t]
\begin{center}
\includegraphics{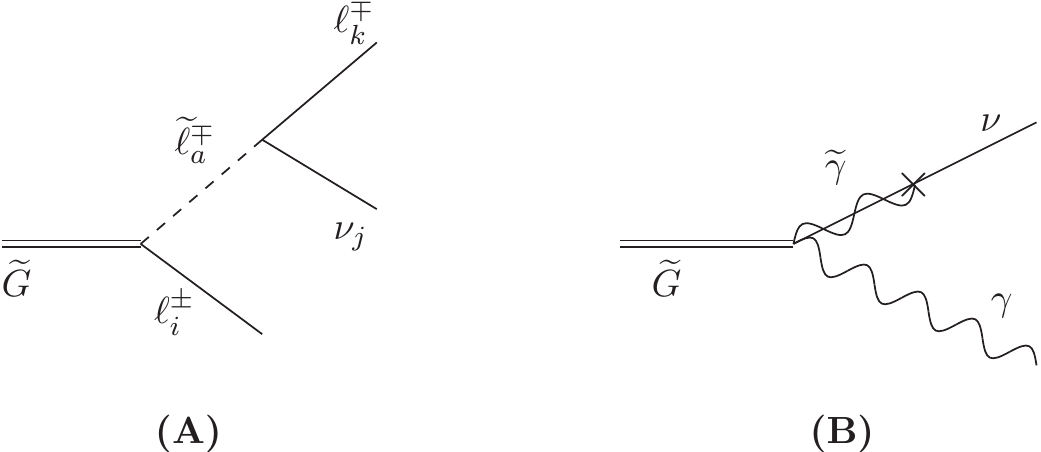}
\caption{Feynman diagrams for $\widetilde G$ decay through {\bf (A)} an off-shell slepton to a three-body final state ($\widetilde G \to \ell_i^\pm  \nu_j \ell_k^\mp $) and {\bf (B)} a two-body radiative process ($\widetilde G \to \nu_\alpha \gamma$).}
\label{fig:grav}
\end{center}
\end{figure}

Once gravitinos are produced with the observed relic abundance, their decay rate is controlled by the strength of the RPV vertex, as well as by the mass of the observable superpartners. The RPV operator considered here allows decays in one of two ways, as shown in the diagrams of \Fig{fig:grav}.  First, through an off-shell slepton, one has $\tilde{G} \rightarrow \nu_j \ell_i^+ \ell_k^-$.  This process is suppressed both by three-body phase space and by the slepton propagator, which gives an additional factor proportional to $(m_{3/2}/\tilde{m})^4$, where $\tilde{m}$ is the slepton mass.  One finds \cite{Buchmuller:2007ui}
\begin{eqnarray}
\tau_{\widetilde G \rightarrow \nu_j \ell_i^+ \ell_k^-} & \simeq & \bL \frac{|\lambda_{i jk}|^2}{3(32)^2\pi^3} \frac{m_{3/2}^3}{m_{\rm Pl}^2}\  F\left(\frac{\tilde{m}}{m_{3/2}}\right) \bR^{-1} \\ \nonumber
& \simeq &  1.0 \times 10^{53}\s  \pL \frac{10^{-4}}{\lambda_{i jk}} \pR^2 \pL \frac{10\mev}{m_{3/2}} \pR^7 \pL \frac{\tilde m}{1 \tev} \pR^4 \,,
\end{eqnarray}
where $m_{\rm Pl} = M_{\rm Pl} /\sqrt{8\pi}= 2.4\times 10^{18}$ GeV is the reduced Planck scale and $F(x) \simeq 1/(30 x^4)$; a more exact expression can been found in \cite{Moreau:2001sr}.

\begin{figure}[t!]
\begin{center}
\includegraphics[width=0.47\textwidth]{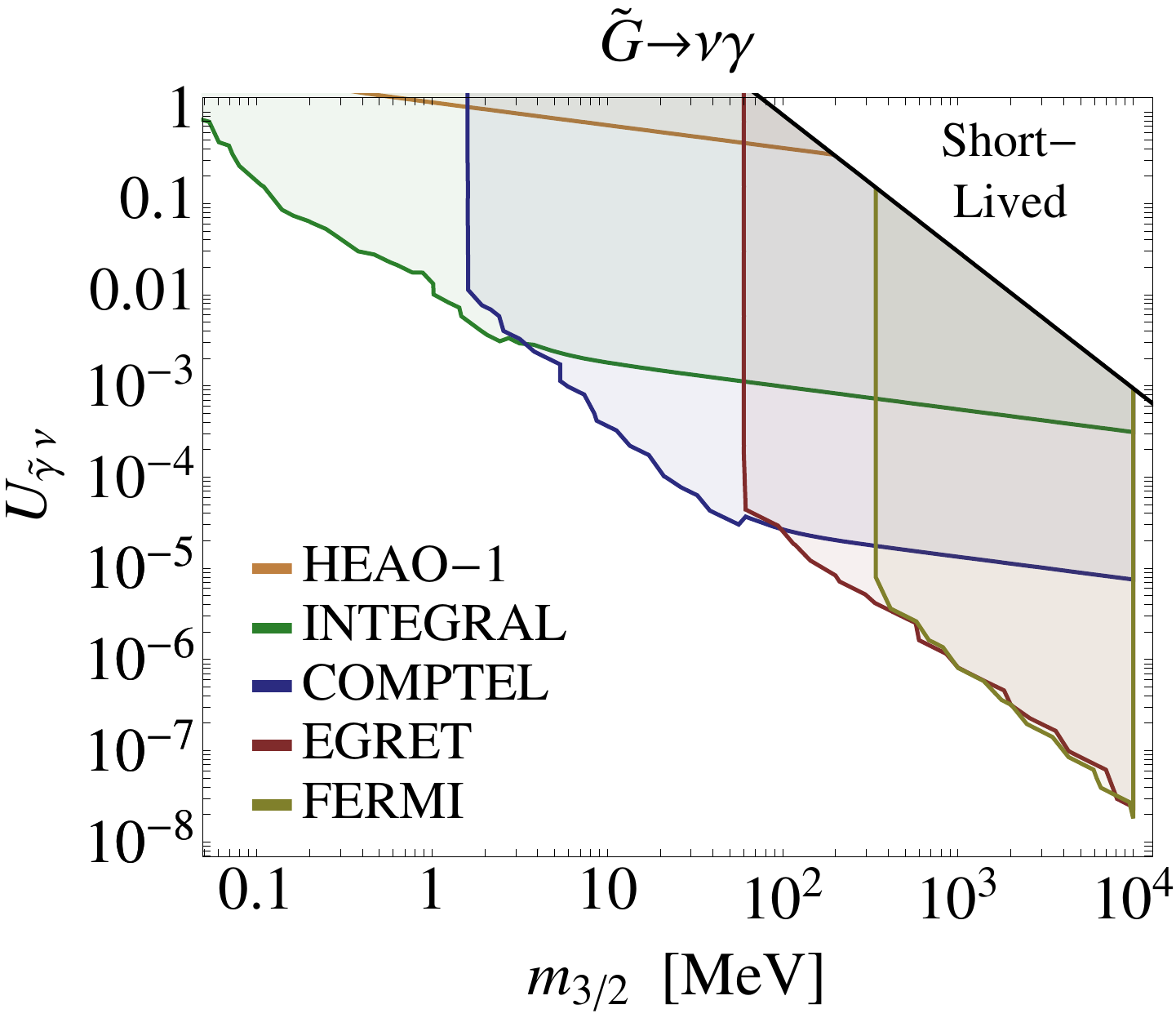}~~~~\includegraphics[width=0.47\textwidth]{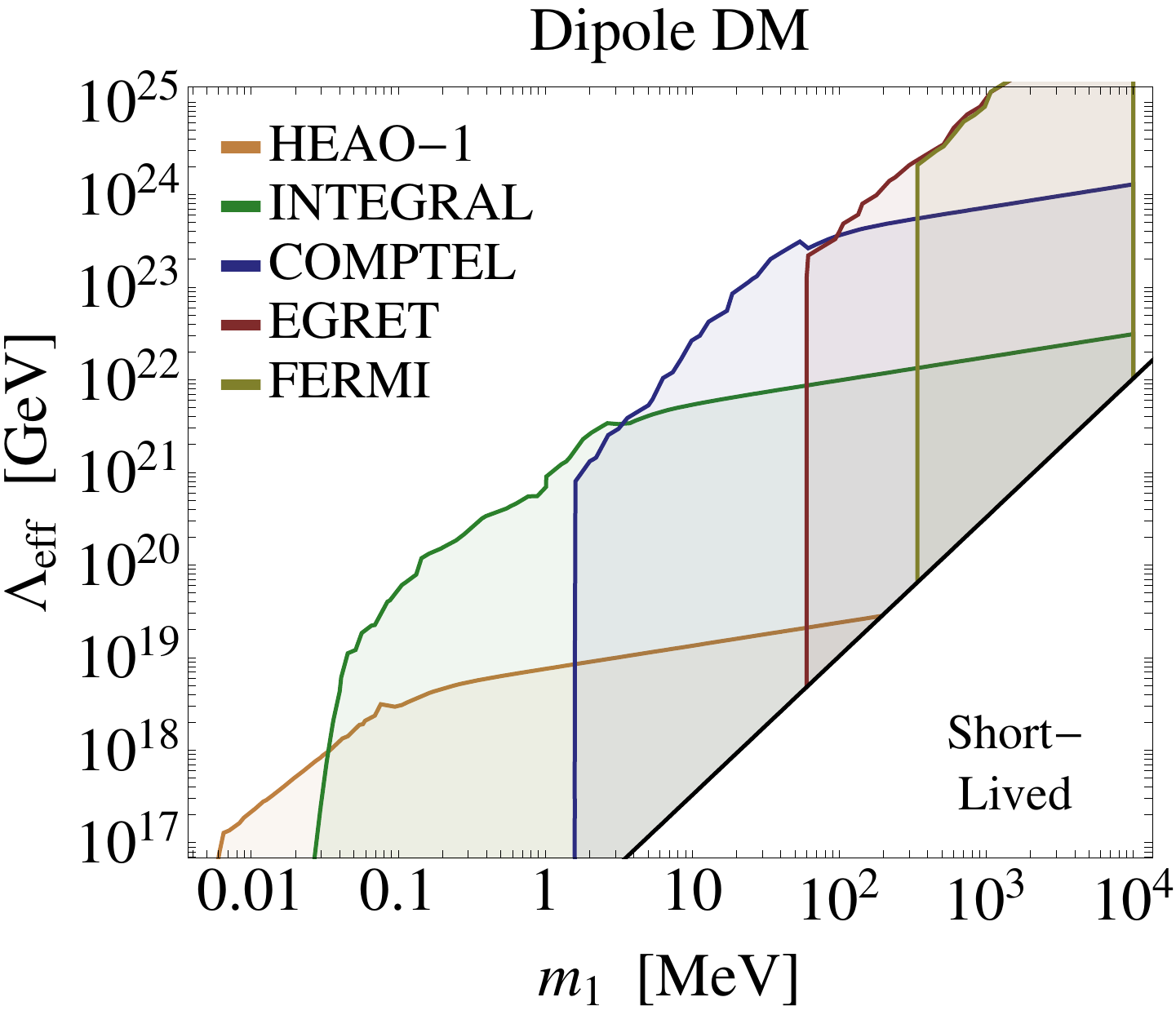}
\caption{{\bf Left:} Constraints on photino-neutrino mixing from RPV gravitino decay.  {\bf Right:} Constraints on the effective cutoff scale for DM with a dipole interaction. Regions as in \Fig{fig:darkphoton}.}
\label{fig:gravitino}
\end{center}
\end{figure}

A second, two-body, decay mode is $\tilde{G} \rightarrow \gamma \nu$,  which usually dominates the decay width~\cite{Buchmuller:2007ui} and gives stronger bounds.  It is induced by a mixing between the photino and the neutrino, $|U_{\tilde \gamma \nu}|$, which occurs if the RPV terms induce a VEV for the sneutrino~\cite{Takayama:2000uz,Buchmuller:2007ui} or via a loop with a charged lepton and slepton.  This gives a gravitino lifetime~\cite{Moroi:1995fs,Takayama:2000uz},
\alg{ \label{grav1}
\tau_{\widetilde G \to \nu \gamma} = \pL \frac1{32\pi} \mL U_{\widetilde \gamma  \nu} \mR^2 \frac{m_{3/2}^3}{m_{\rm Pl}^2} \pR^{-1}
 \simeq 3.8 \times 10^{28}\s \pL \frac{10\mev}{m_{3/2}}\pR^3 \pL \frac{10^{-4}}{U_{\widetilde \gamma  \nu}}\pR^2\,.
}
In the left panel of Fig.~\ref{fig:gravitino} we show the constraints on the photino-neutrino mixing angle as a function of the gravitino mass.  In deriving the bound we require that the gravitino has the observed DM relic abundance.   We do not show limits from BBN as those depend strongly on the dominant production mechanism and hence on the re-heat temperature and the spectrum of the superpartners~\cite{Hall:2013uga}. 

 \subsection{Dipole DM}

The dipole operator $ \lambda \bar \chi_2 \sigma^{\mu \nu}\chi_1 F_{\mu \nu}/\Lambda$ (with $\sigma^{\mu \nu}=i\bL \gamma^\mu,\gamma^\nu \bR$) induces $\chi_1 \rightarrow \chi_2 \gamma$, where $\chi_{1,2}$ are Dirac fermions.  The lifetime is 
\alg{
\tau_{\rm dipole} &= \bL  \frac{m_1^3}{2\pi \Lambda_{\rm eff}^2}\left(1-\frac{m_2^2}{m_1^2}\right)^3 \bR^{-1}
\\ & \simeq  4.1 \tenx{20} \s \pL \frac{10\mev}{m_1} \pR^3 \pL \frac{\Lambda_{\rm eff}}{10^{19}\gev} \pR^2\,,
}
with $\Lambda_{\rm eff} = \Lambda/\lambda$, the effective cutoff scale of the theory.  
The outgoing photon has an energy $E_\gamma = \pL m_1^2 - m_2^2 \pR/2m_1$. 
 In the right panel of \Fig{fig:gravitino}, we show the limits on $\Lambda_{\rm eff}$ versus the $\chi_1$ mass, $m_1$.  
Since the effective operator that controls the decay is dimension 5 and not higher-dimensional, the limits are exceptionally strong, constraining the effective cutoff scale to be very high (or conversely, the corresponding coupling  to be small,  $\lambda\ll1$).  An approximate symmetry in the UV may be required to protect these decays.

 \subsection{Dark (Pseudo-) Scalars}

 \begin{figure}[t!]
\begin{center}
\includegraphics[width=0.47\textwidth]{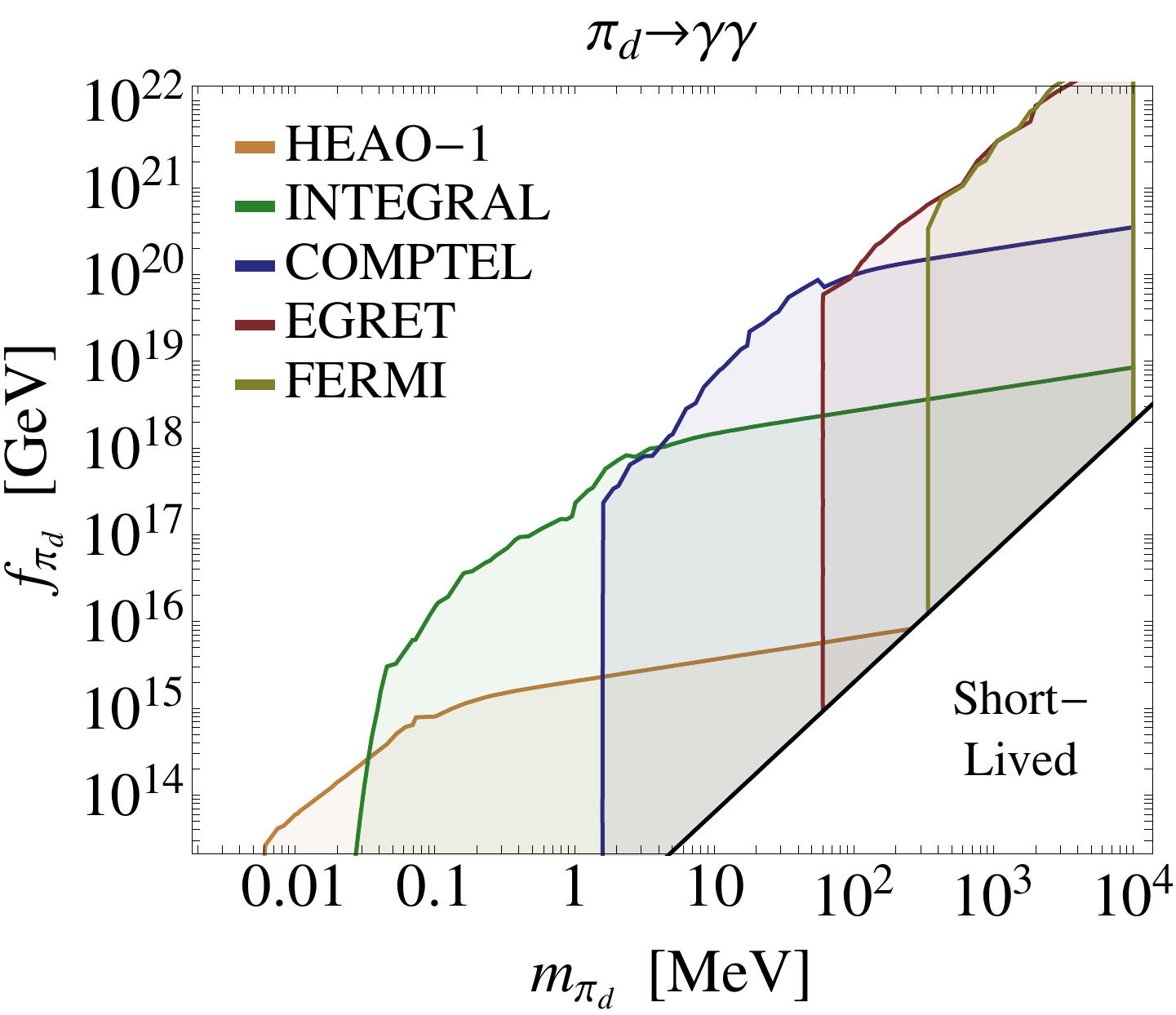}~~~~~\includegraphics[width=0.48\textwidth]{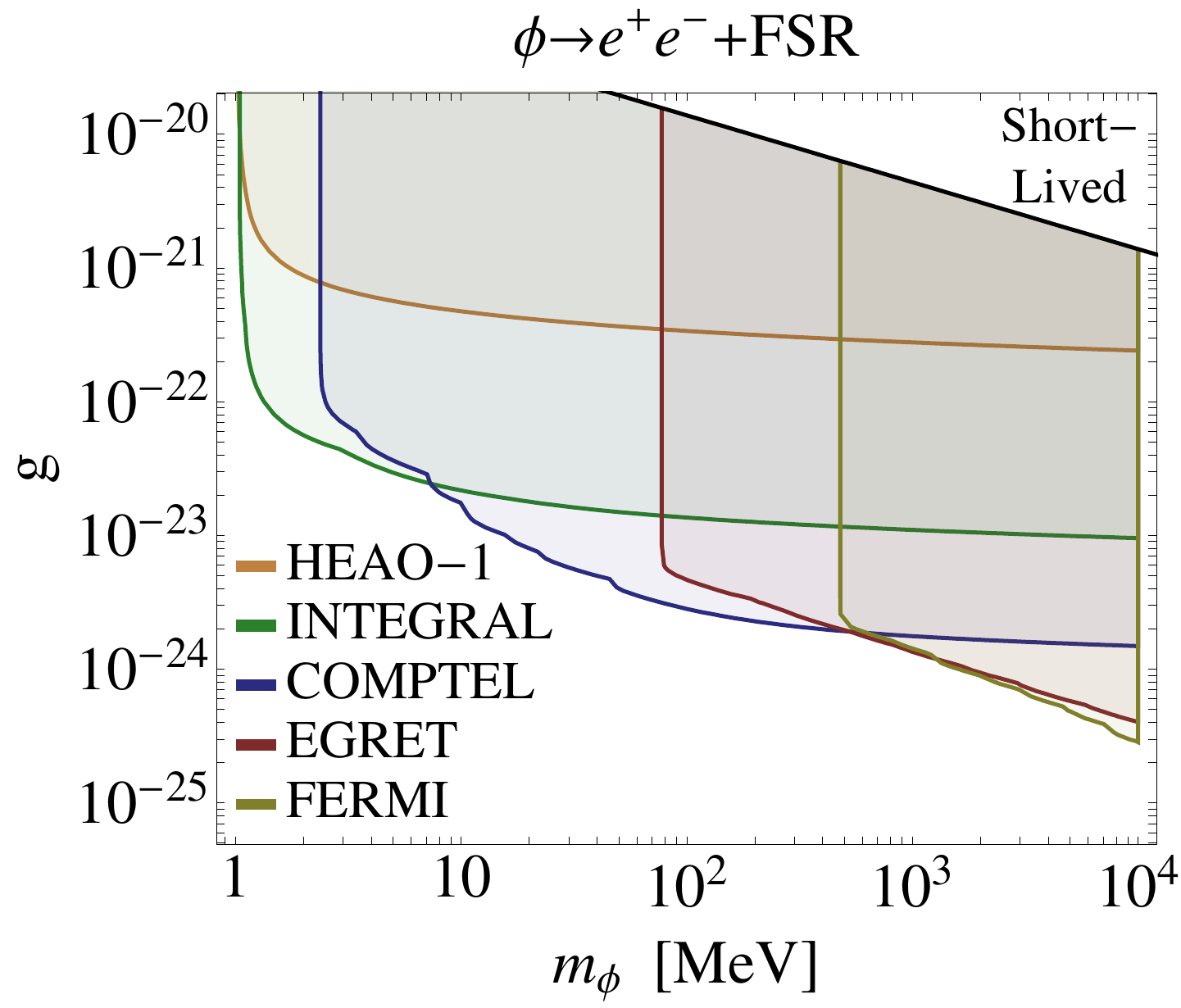}
\caption{Constraints on the decay constant $f_{\pi_d}$ for a dark pseudoscalar  decaying to diphotons {\bf (left)} and the limits on the coupling of a hidden scalar in the case where it decays to $e^+e^-$  {\bf (right)}.  Regions as in \Fig{fig:darkphoton}. }
\label{fig:pseudoscalar}
\end{center}
\end{figure}

As a final model for light DM, we consider two-body decays of diphotons or charged particles.   
 If DM is a pseudoscalar decaying to two photons, its lifetime is~\cite{Langacker:2010zza}
\alg{
\tau_{\pi_d \to \gamma \gamma} &\simeq \pL \frac{\alpha_{\rm EM}^2 m_{\pi_d}^3}{288\pi^3 f_{\pi_d}^2}  \pR^{-1} \simeq  1.1 \tenx{20} \s \pL\frac{10\mev}{m_{\pi_d}} \pR^3 \pL\frac{f_{\pi_d}}{10^{15}\gev} \pR^2 \,.
}
Here $f_{\pi_d}$ is the decay constant in the hidden sector, which we assume is Abelian. This decay produces a spectral line at an energy $m_{\pi_d}/2$. 
 We show the constraint in the left panel of Fig.~\ref{fig:pseudoscalar}, from which it is clear that the scale of $f_{\pi_d}$ needs to be very high.

If DM is a scalar that decays to charged particles that produce photons through FSR, {\it e.g.},~$\phi \to e^+ e^-$, the lifetime is 
\beq
\tau_{\phi \to e^+e^-} = \bL \frac{g^2 m_{\pi_d}}{4\pi} \pL 1 - 4 \frac{m_e^2}{m_{\rm DM}^2} \pR^{3/2} \bR^{-1}  \simeq  8.3 \times 10^{18}\s \frac{10\mev}{m_{\phi}} \pL\frac{10^{-20}}{g_a}\pR^2\,.
\eeq
The spectrum is bounded by the energies $0<E_\gamma<m_{\phi}/2$.   The constraints on the coupling $g$ are shown in the right panel of Fig.~\ref{fig:pseudoscalar}.  As is apparent, tiny couplings are required for such DM to agree with observations.

\section{Model-Independent Bounds and Spectra}
\label{sec:spectra}

In the previous section, we presented limits on specific model parameters.  In this section, we fill in some of the details of the analysis there, and show bounds in terms of the lifetime only, making the constraints ``model-independent.''
Despite the wide variety of possible decays that produce a photon signal, 
there are very few distinct event topologies of interest: 
\begin{itemize}
\item Two- or three-body decays, with or without FSR.
\item Two-body cascade decays, where one or both of the decay products themselves subsequently decay to photons or charged particles.
\end{itemize}
In the limit of small outgoing particle masses, the differential decay width at low energies for each of these topologies may be written as a function of the total width, the photon energy, $E_i$, and the mass of the outgoing particle, $m_i$. 
We will use the small parameters 
\beq  \label{small}
\nu_i=\frac{m_i}{m_{\rm DM}}\, , \qquad \qquad \lambda_i=\frac{E_i}{m_{\rm DM}}\,,
\eeq
to expand our results. 

When relevant in the model-independent bounds below, we only consider photons and electrons as SM  final states.   Typically these bounds will weaken moderately as new decay channels  to additional charged or unstable heavier particles open up.  One exception, however, is for the case where the decay products include $\pi^0$'s which consequently decay to photons.  In such a case, a significant improvement in the limits is expected due to the sharp spectral feature.

\subsection{Two-Body Decays Involving a Photon}
\label{subsec:twobodyphoton}

\begin{figure}[t]
\begin{center}
\includegraphics[width=0.8\textwidth]{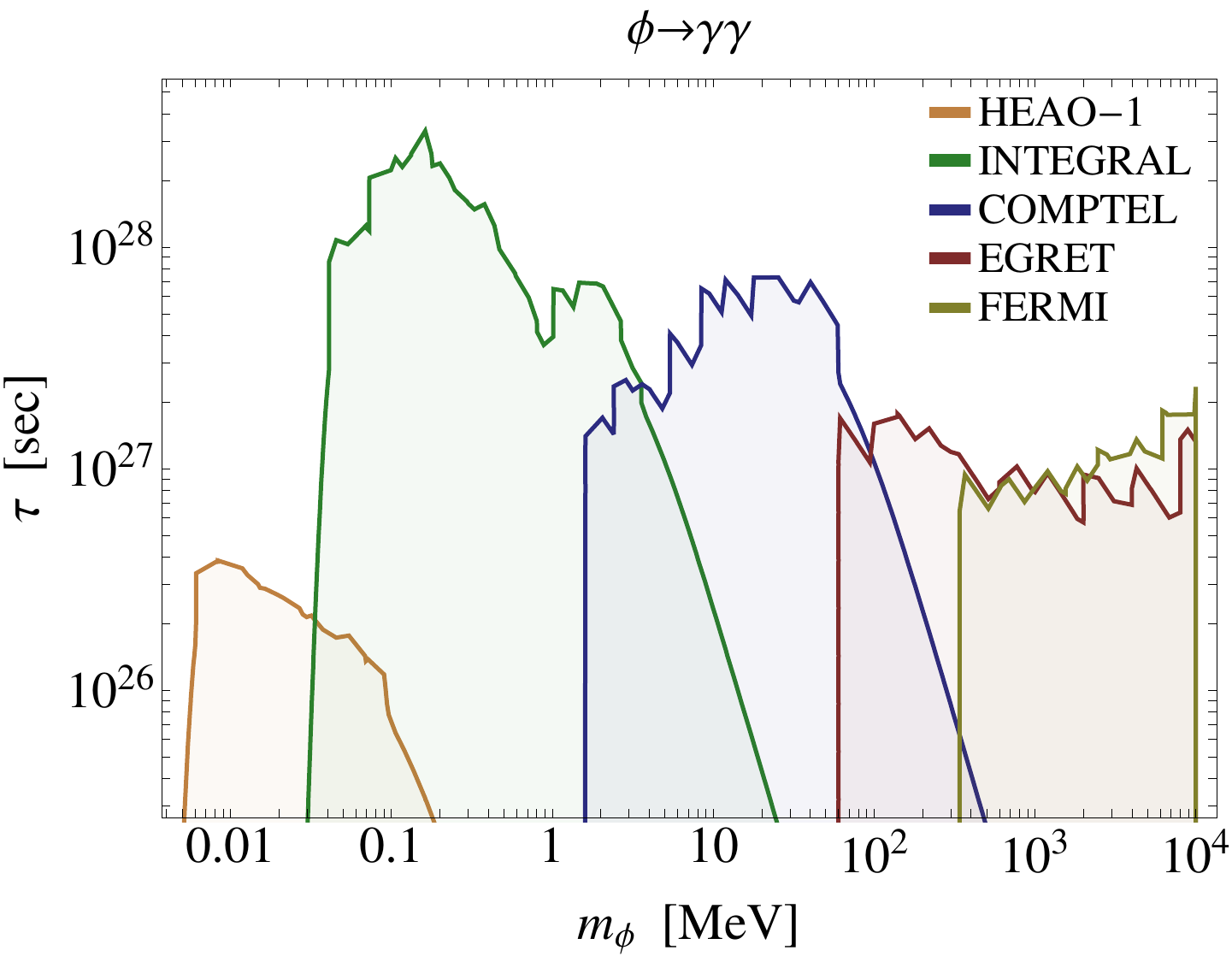}
\caption{Bounds on the lifetime of a scalar DM, $\phi$,  decaying to two photons.   Regions as in \Fig{fig:darkphoton}.}
\label{fig:line}
\end{center}
\end{figure}

We first consider  two-body decays of DM directly to a photon and a neutral particle, or to two photons. Models that give line-like features include a hidden photino decaying to a gravitino and a photon via kinetic mixing, as discussed in \Sec{sec:hidgaugino}. There are, 
of course, a profusion of other model-building possibilities that produce a monochromatic photon. These decays can produce one or two 
monochromatic photons with differential width,
\beq \label{eq:line}
\frac{dN_{\rm two-body} }{dE_\gamma}= \cbL \begin{array}{cl} \delta(1- \nu_2^2- 2 \lambda_\gamma) & {\rm ~(1~photon)} \\ 
2  \delta(1- 2 \lambda_\gamma) & {\rm ~(2~photons)} \end{array} \right. \,.
\eeq
Here $\nu_2\equiv m_2/m_{\rm DM}$ refers to the mass of the outgoing decay partner, in the case of a single photon. 
The constraints on the lifetime for the decay to two photons are shown in Fig.~\ref{fig:line}.  

\subsection{Two-Body Decays with FSR}
\label{subsec:twobodyfsr}

\begin{figure}[t]
\begin{center}
\includegraphics[width=.47\textwidth]{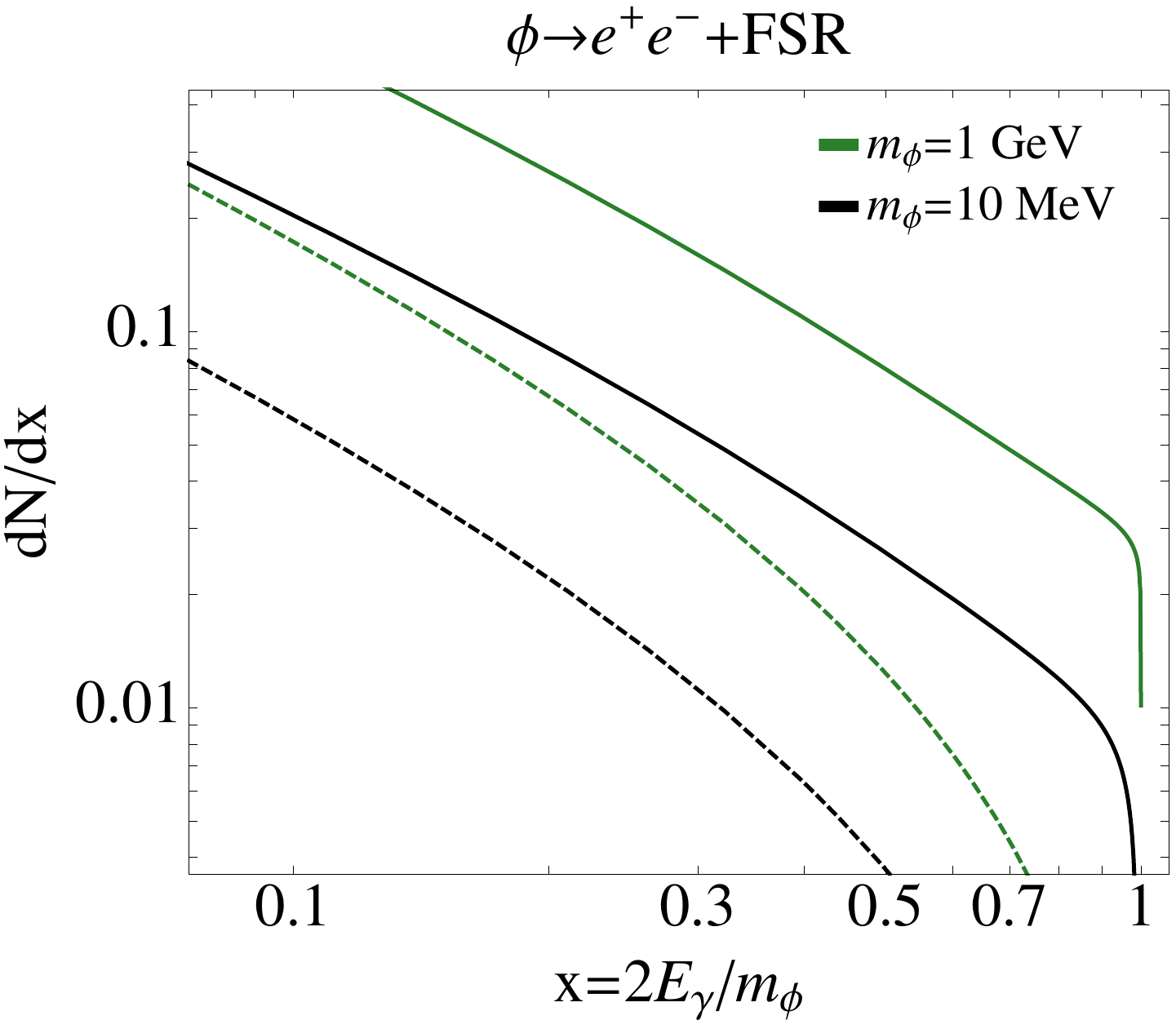}~~~~ \includegraphics[width=.472\textwidth]{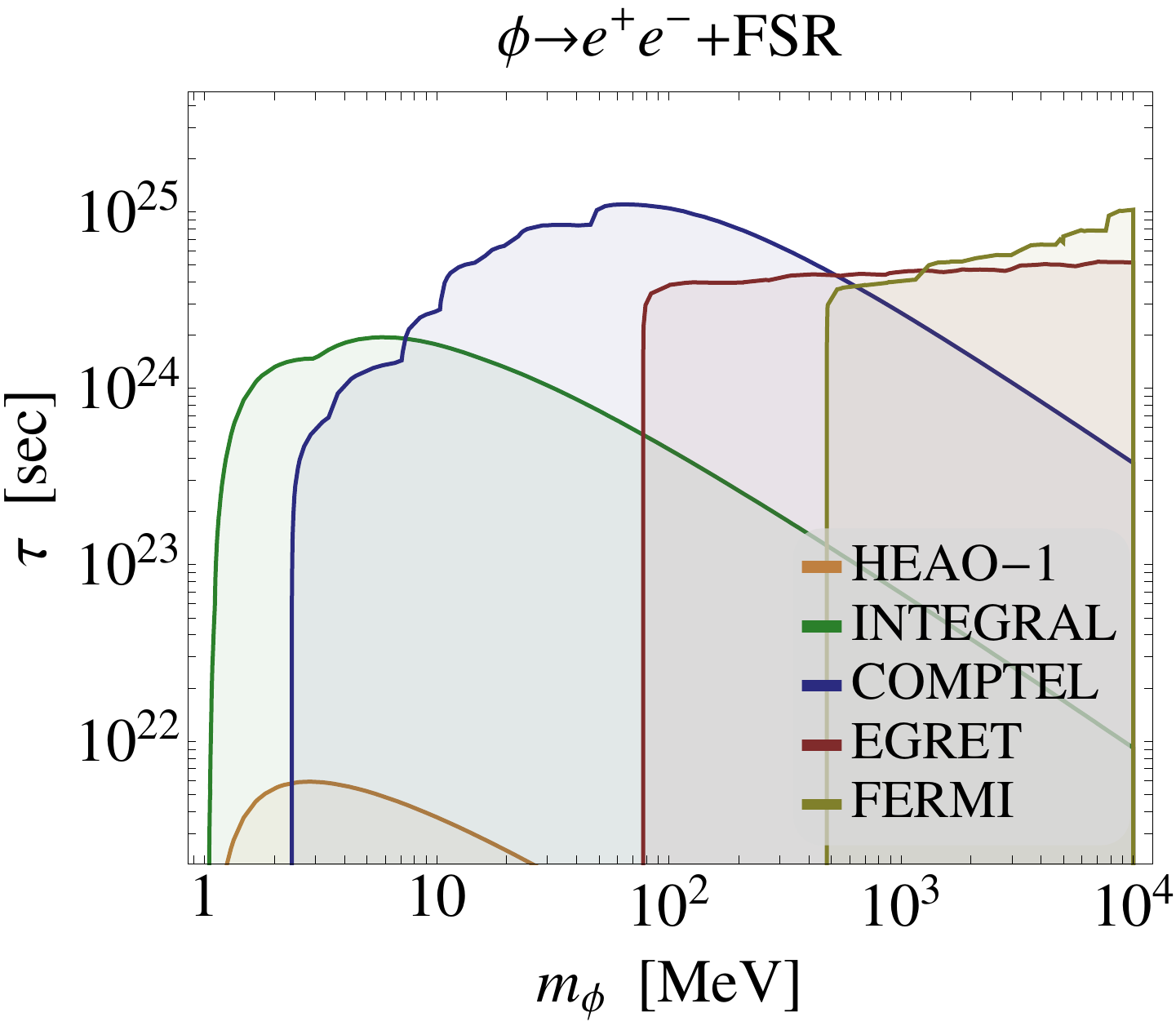}
\caption{{\bf Left}: Photon spectra from DM decay to $e^+ e^-$, emitting final state radiation, as a function of $x=2 E_\gamma/m_{\rm DM}$. The spectrum of decays with galactic photons only is shown as the solid line, while the redshifted extragalactic spectrum is shown with dashed lines (see text for details).  {\bf Right}:  Bounds on the DM decay lifetime for this process, with regions as in \Fig{fig:darkphoton}.}
\label{fig:lineFSR}
\end{center}
\end{figure}

Two-body decays to charged particles produce photons through FSR.  The differential width to photons is approximately given by integrating a $\delta$-function with the Altarelli-Parisi splitting function, as shown in Eq.~\eqref{apsf}, to give
\alg{ \label{eq:phieegam}
\frac{dN_{\phi \to e^+e^- \gamma}}{dE_\gamma} &\simeq \frac{2\alpha_{\rm EM}}{\pi E_\gamma} \bL 1 - 2\lambda_\gamma+ \pL 1-2\lambda_\gamma + 2\lambda_\gamma^2\pR \ln \pL \frac{1-2\lambda_\gamma}{\nu_e^2} \pR \bR,
}
where the spectrum is bounded by the energies $0<E_\gamma<m_\phi/2$. We use the exact calculation of the three-body final state for the spectra and the exclusion regions in \Fig{fig:lineFSR}. In this figure, we show the dimensionless galactic photon spectrum 
\beq
\frac{dN}{dx}=\frac{m_1}2 \frac{dN}{dE}
\eeq
as well as the redshifted extragalactic spectrum $dN_{\rm \gamma,eg}/dx$ (dashed lines). The extragalactic spectrum is calculated by performing the integral in Eq.~(\ref{eq:redshifted})
\alg{
\frac{dN_{\rm \gamma,eg}}{dx} = \frac{\int_0^\infty dz ~ \frac{dN}{dx(z)} ~\bL  (1+z)^3 +\kappa \bR^{-1/2}}{\int_0^\infty dz ~\bL  (1+z)^3 +\kappa \bR^{-1/2}},
}
normalized such that the total number of photons for $0<x<1$ is equivalent for galactic and extragalactic photons. 

As described above, this decay naturally arises if the DM is a light scalar. Furthermore, the decay to two leptons is a popular toy model that parameterizes possible DM decay and annihilation. The bounds for this case are shown on the right of Fig~\ref{fig:lineFSR}.  As expected,  they are a few orders of magnitude weaker than the bounds from the monochromatic decay shown in the previous subsection.

\subsection{Two-Body Cascade Decays}
\label{subsec:twobodycascade}

\begin{figure}[t!]
\begin{center}
\includegraphics[width=.47\textwidth]{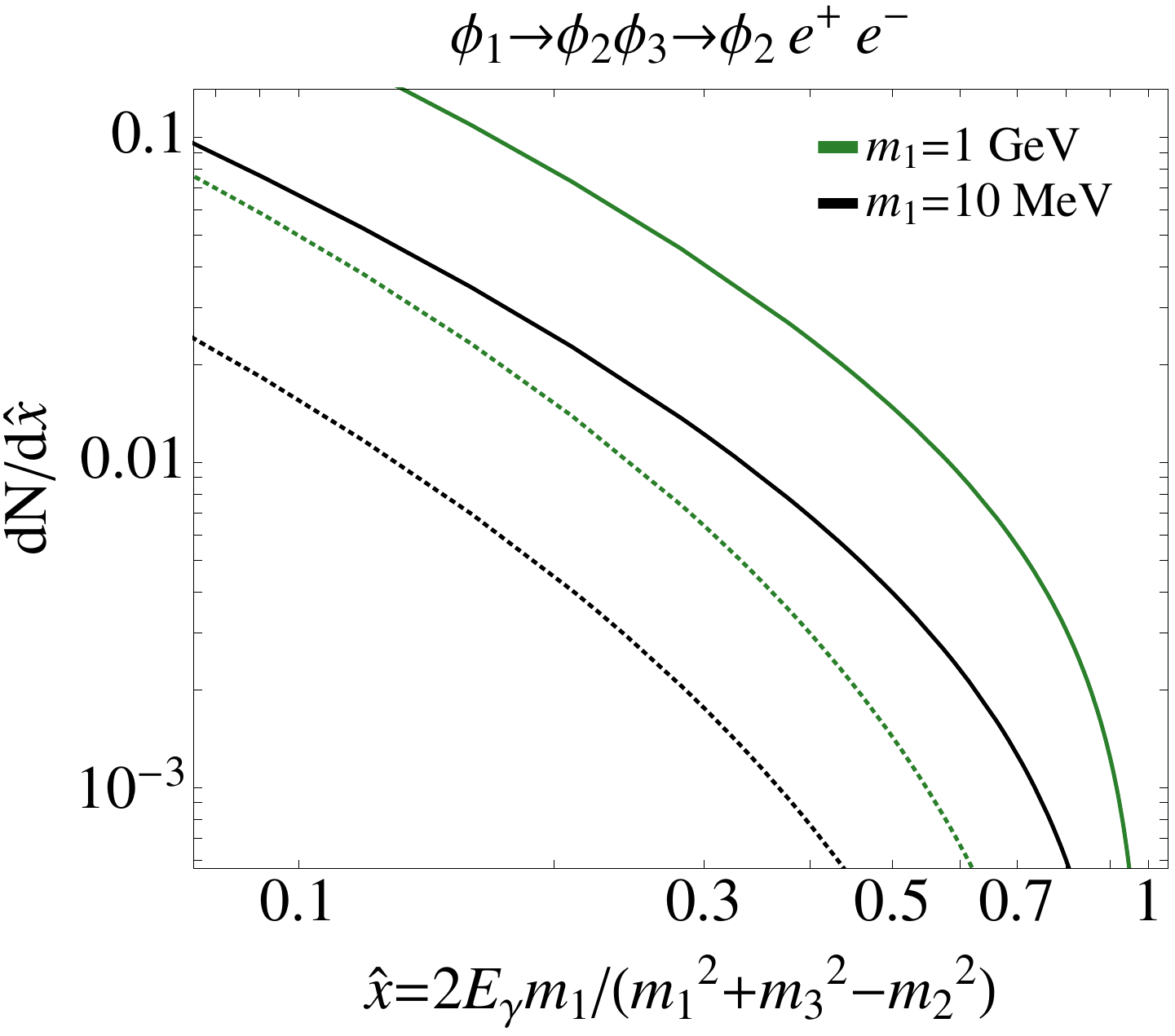}~~~~~ \includegraphics[width=.472\textwidth]{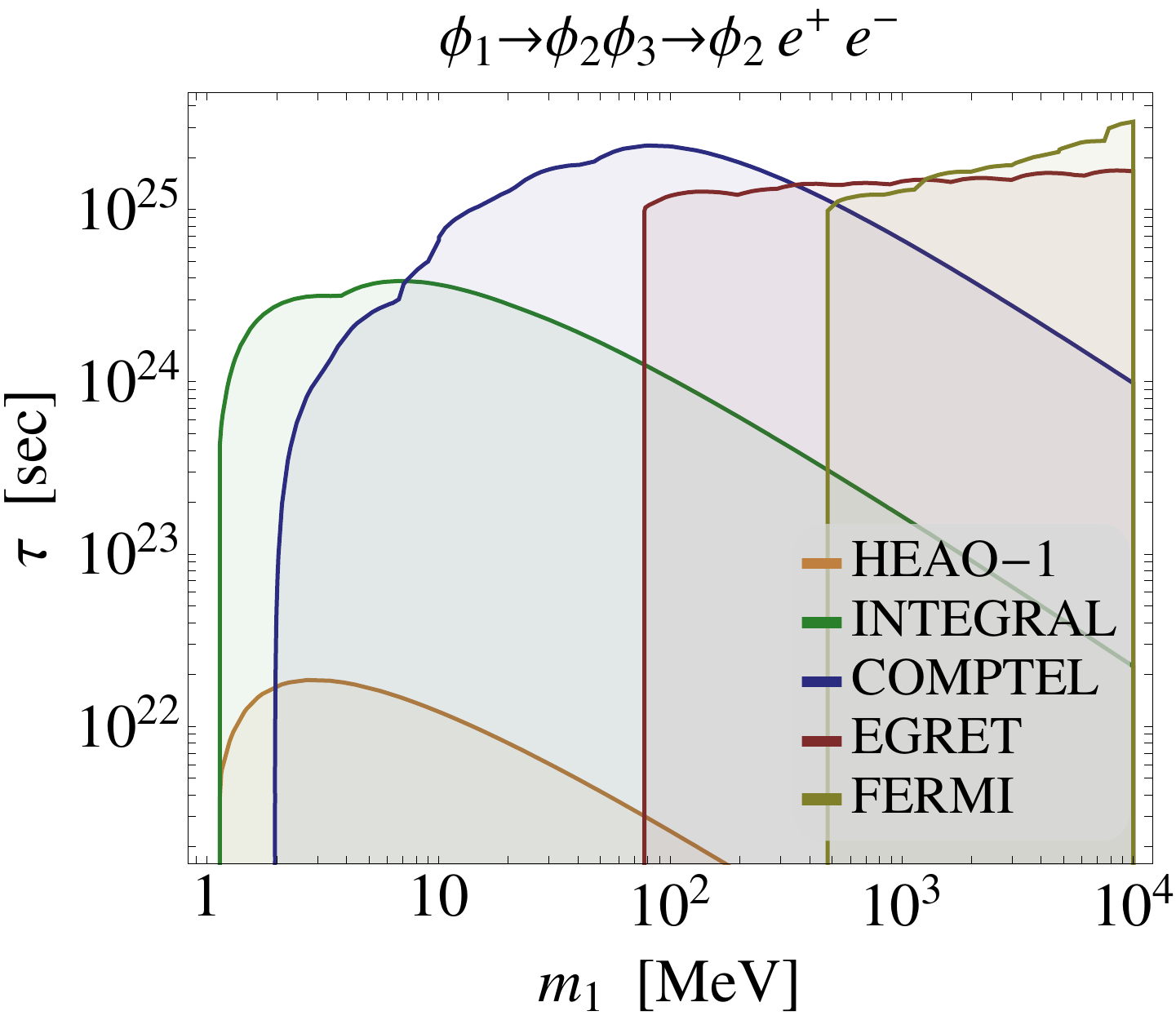}
\caption{{\bf Left}: Photon spectra versus $x=2E_\gamma/m_{\rm DM}$ for DM decay to two neutral particles, where one of the neutral particles subsequently decays to $e^+ e^-$, emitting final state radiation.  The lines are as in \Fig{fig:lineFSR}. {\bf Right}: Bounds on the DM decay lifetime for this process.  Regions are as in~\Fig{fig:darkphoton}. }
\label{fig:lineboost}
\end{center}
\end{figure}

We next consider the case of DM decay to a pair of neutral particles, one of which subsequently decays to $e^+ e^-$:  $\phi_1 \to \phi_2 \phi_3 \to \phi_2 \ell^+ \ell^-$.  An example for a decay of this type was presented in \Sec{sec:hidgaugino} for the hidden photino model, where the hidden photino decays to a gravitino and hidden photon, which then subsequently decays to charged leptons: $\widetilde \gamma_d \to \widetilde G   \gamma_d \to \widetilde G \ell^+ \ell^-$. We derive the photon spectrum from these cascade decays from~\cite{Mardon:2009rc}.  The spectrum for FSR resulting from a single boosted lepton is
\begin{multline} \label{eq:boost}
\frac{dN}{dE_\gamma}= \frac{2\alpha_{\rm EM}}{ \pi m_1 \widehat x} \Bigg\{ \bL -1+\ln \pL \frac{m_3^2}{m_\ell^2} \pR \bR \pL 2- \widehat x- \widehat x^2+2 \widehat x \ln \widehat x \pR + \pL \frac{\pi^2}3-1 \pR \widehat x+\widehat x^2 +\\ 2\widehat x \ln \widehat x + \pL 2-\widehat x-\widehat x^2 \pR \ln \pL 1-\widehat x \pR -2 \widehat x {\rm Li}_2 \pL \widehat x \pR \Bigg\},
\end{multline}
where  $\widehat x= 2m_1E_\gamma/\pL m_1^2+m_3^2-m_2^2 \pR$.   This spectrum, under the assumption of  $m_3= 0.9 m_1 $ and $m_2 = 0.01m_1$, is shown on the left of Fig.~\ref{fig:lineboost} where the galactic (solid lines) and redshifted extragalactic (dashed lines) contributions are shown.    As can be seen, \Eq{eq:boost} does not have a precise cutoff at $E_\gamma=m_1/2$. However, as noted in~\cite{Mardon:2009rc}, the number of unphysical photons produced with $E_\gamma >m_1/2$ is second order in the expansion parameters and the effect of this error on the bounds is negligible.

The constraints on the lifetime of the decaying particle are shown on the right of Fig.~\ref{fig:lineboost}, (with similar assumptions on $m_{2,3}$ as made in the left panel).  These constraints are comparable to those on two-body + FSR models, and are considerably less constraining than those with monochromatic photons. 

\subsection{Three-Body Decays with FSR}
\label{subsec:threebodyfsr}

\begin{figure}[t!]
\begin{center}
\includegraphics[width=.47\textwidth]{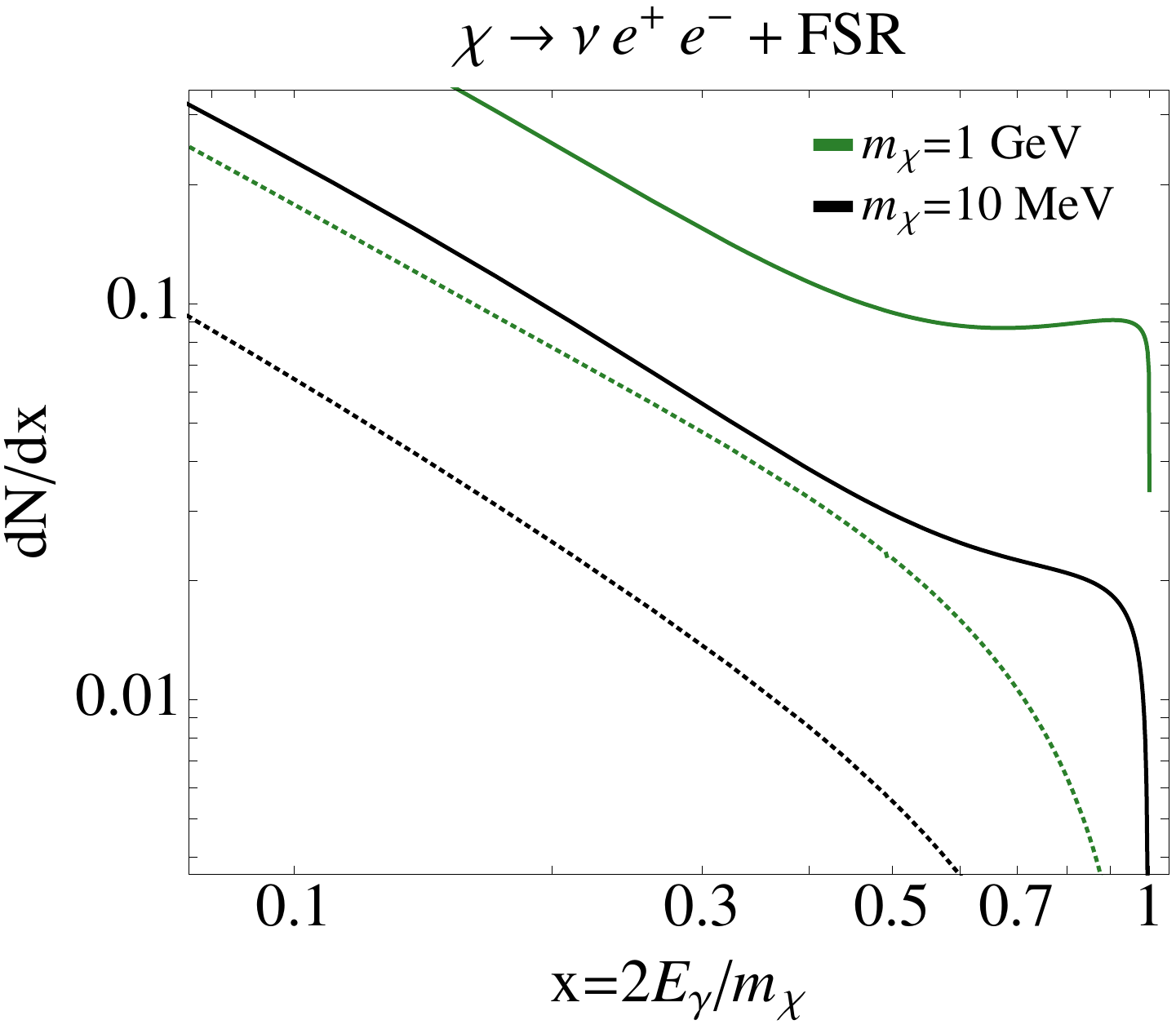}~~~~~ \includegraphics[width=.474\textwidth]{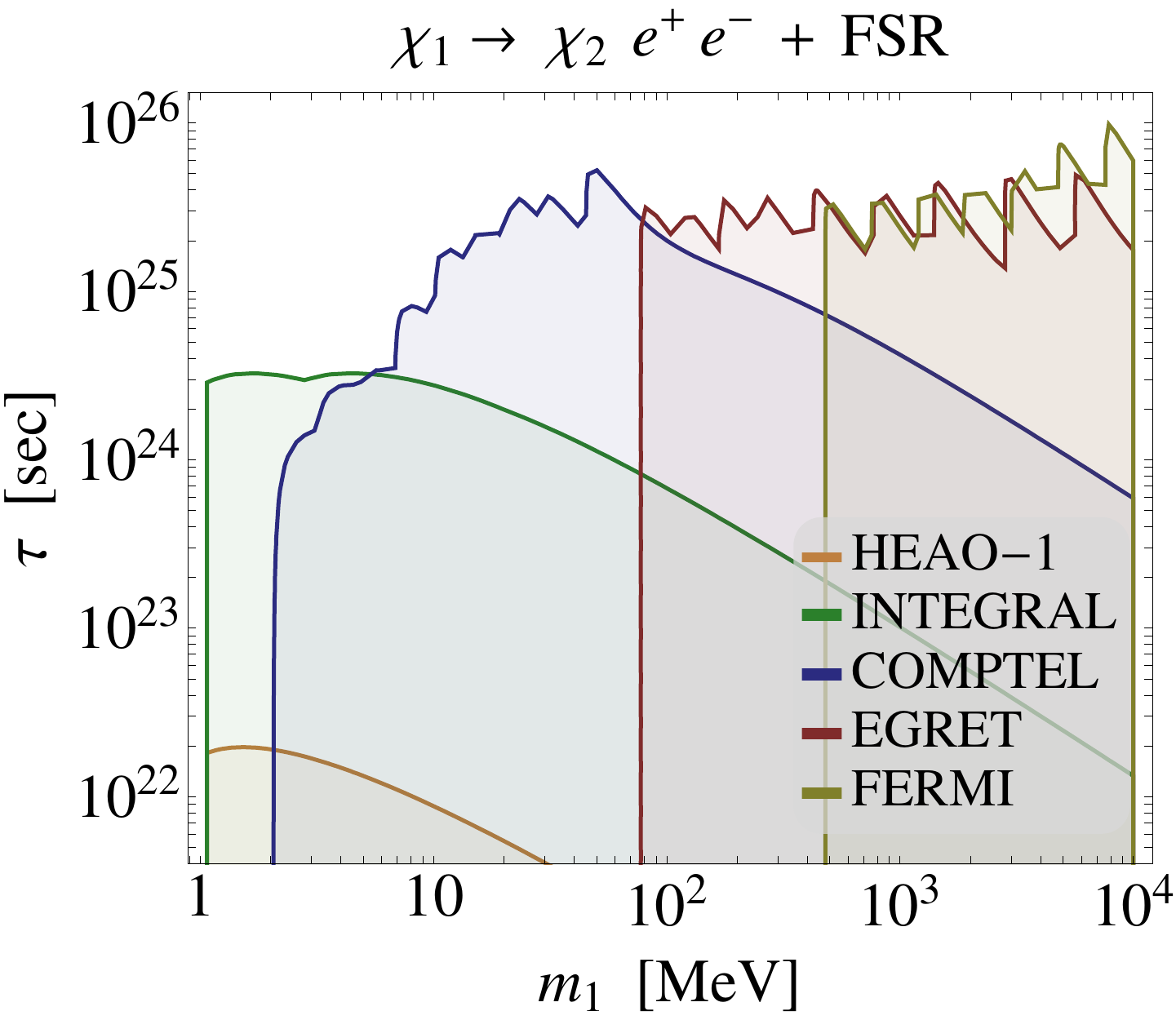}
\caption{{\bf Left}: Photon spectra versus $x=2E_\gamma/m_{\rm DM}$ for DM decay to $e^+ e^- \nu$, emitting final state radiation.  The lines represent the galactic (solid) and extragalactic (dashed) spectra. {\bf Right}: Bounds on the DM decay lifetime for this process. Regions as in \Fig{fig:darkphoton}.}
\label{fig:3w}
\end{center}
\end{figure}

Next we examine three-body DM decays, where the DM decays to a pair of charged particles plus a neutral particle.  Our formula was specifically derived for the case of Weak decays of a sterile neutrino, $\nu_s \rightarrow \nu e^+ e^-$ (as we discussed in \Sec{sec:sterileneutrino}), though only minor changes result for a more generic decay $\phi_1 \rightarrow \phi_2 e^+ e^-$. 

The differential width of a fermionic DM decaying  to $e^+ e^- \nu$ via weak processes and including FSR  is,
\beq  \label{nusterile3body2}
\frac{d N_{DM,{\rm FSR}}}{d E_\gamma} \simeq \frac{2 \alpha_{\rm EM}}{ \pi E_\gamma} \log \pL \frac{1-2\lambda_\gamma}{\nu_e^2} \pR \bL  1-\frac{11}3\lambda_\gamma + 10 \lambda_\gamma^2 +\frac{ \lambda_\gamma \pL 1+4\sin^2 \theta_W \pR \pL 1-6\lambda_\gamma \pR}{12 c_\alpha}  + \cdots \bR\,.
\eeq
Here we neglect both the neutrino and the electron masses and ``...'' stands for higher-order terms in $\nu_e$.    For the case of a decay process mediated by a heavy neutral scalar particle, the above remains the same with the omission of the last term.

The spectrum for the above is plotted on the left of Fig.~\ref{fig:3w} where, as before,  the galactic (solid lines) and redshifted extragalactic (dashed lines) contributions are shown.  The constraints on the lifetime are shown on the right of Fig.~\ref{fig:3w}.  We find the bounds to be similar in magnitude to the two-body + FSR case, however sensitivity to the endpoint feature in the spectrum is apparent and results in the wiggles displayed in the figure.

\subsection{Three-Body Decays Involving Photons}
\label{subsec:threebodyphoton}

Three body decays such as $\phi_1 \rightarrow \phi_2 \gamma \gamma$ are also possible.  We remain agnostic about the UV completion and do not embed this interaction in any of the theories above.  Nonetheless, we include it here for completeness.

To obtain bounds, we assume that this decay is induced by the higher-dimensional operator $\cO=\frac\beta{4\Lambda^2} \phi_1\phi_2F_{\mu \nu} F^{\mu \nu}$. We have,
\alg{ \label{eq:phiphigamgam}
\frac{dN_{\phi_1 \to \phi_2 \gamma \gamma}}{dE_\gamma} &= \frac{128 E_\gamma^3}{m_1^4} \frac{ \pL 1- \frac{\nu_2^2}{1-2\lambda_\gamma} \pR^3 }{1+28\nu^2\pL1-\nu_2^4 \pR - \nu_2^8 + 12 \nu_2^2 \pL 1+ 3 \nu_2^2 +\nu_2^4 \pR \ln \nu_2^2 }.
}
We see here that the width is exponentially sensitive to the energy in the limit $\nu_2 \to 0$, which means that the photons from this decay are preferentially grouped near the DM mass.  Consequently, for a given $m_\chi$, the constraint arises from a single bin in a given experiment. 

\begin{figure}[t]
\begin{center}
\includegraphics[width=.46\textwidth]{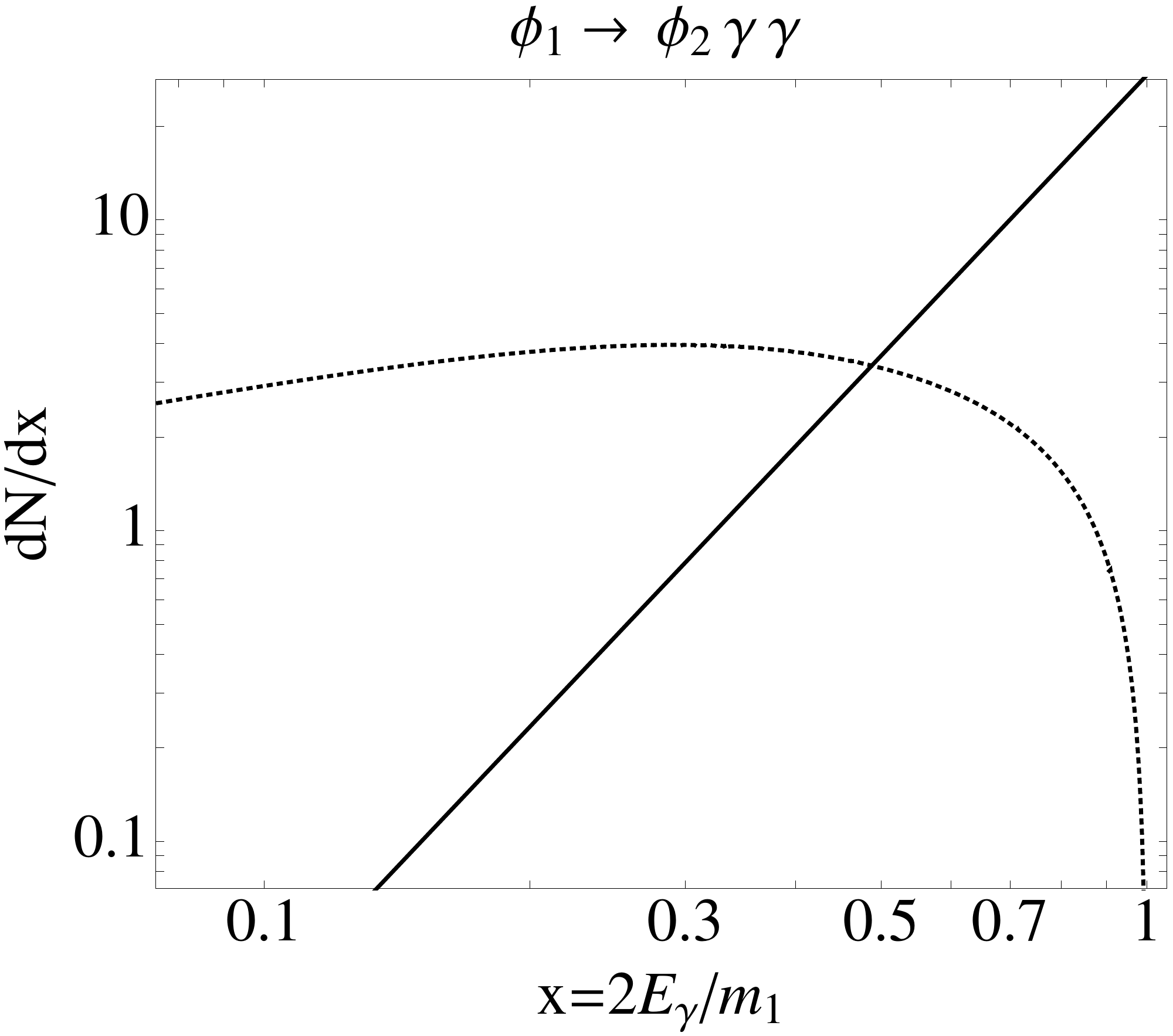}~~~~~ \includegraphics[width=.478\textwidth]{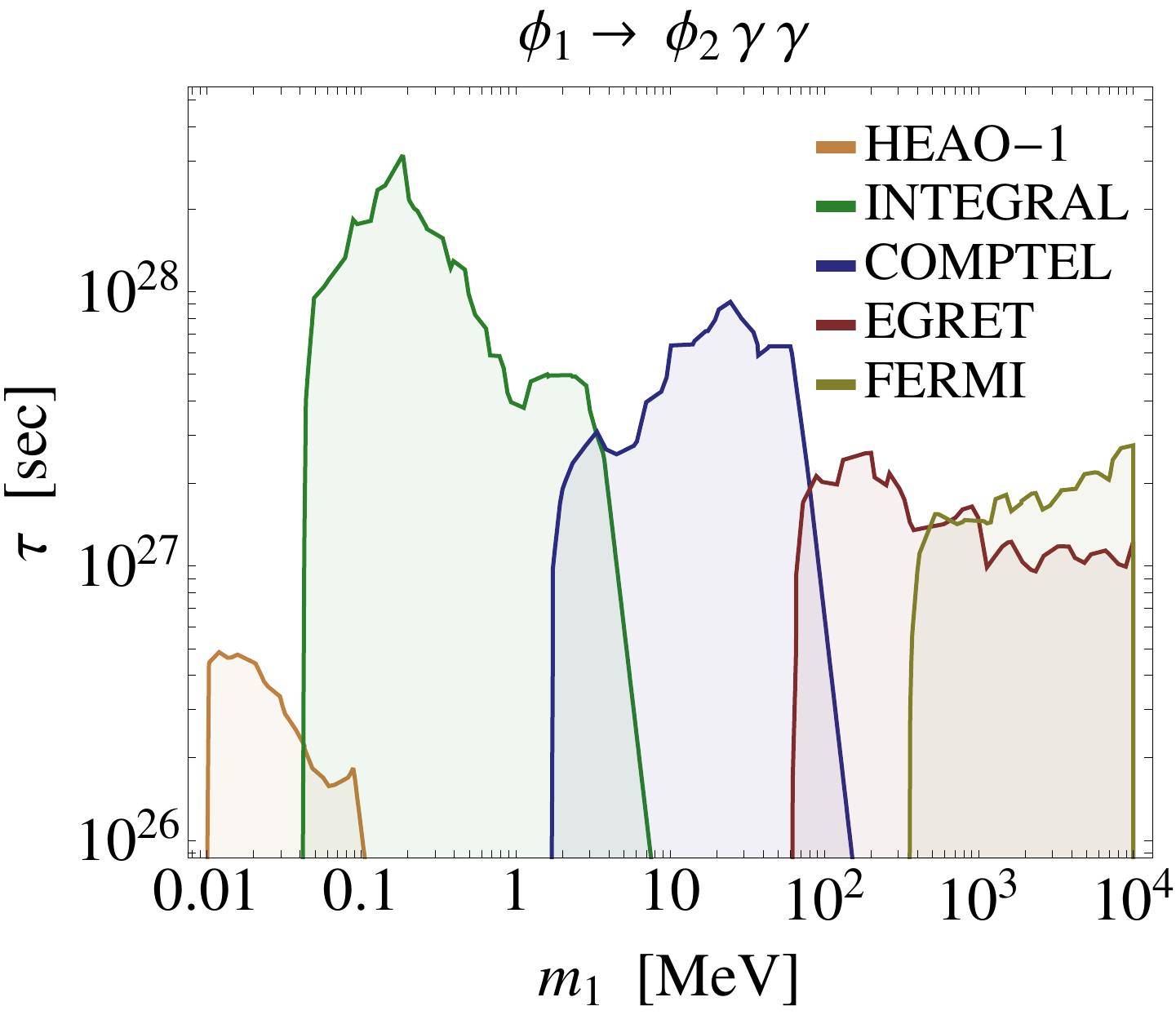}
\caption{{\bf Left}: Photon spectra versus $x=2E_\gamma/m_{\rm DM}$ for DM three-body decay $\phi_1 \rightarrow \phi_2 \gamma \gamma$.  The lines represent the galactic (solid) and redshifted extragalactic (dashed) spectra. {\bf Right}: Bounds on the DM decay lifetime for this process, with regions as  explained in \Fig{fig:darkphoton}.}
\label{fig:3gam}
\end{center}
\end{figure}

We display the spectrum and constraint on the lifetime in Fig.~\ref{fig:3gam}, with the assumption $m_2=0$. In this limit, the differential spectrum is the same regardless of $m_1$.    As expected, these bounds compare favorably to the monochromatic photon lines.


\section{Annihilating Light Dark Matter}
\label{sec:annihilating}

\begin{figure}[t]
\begin{center}
\includegraphics[width=0.8\textwidth]{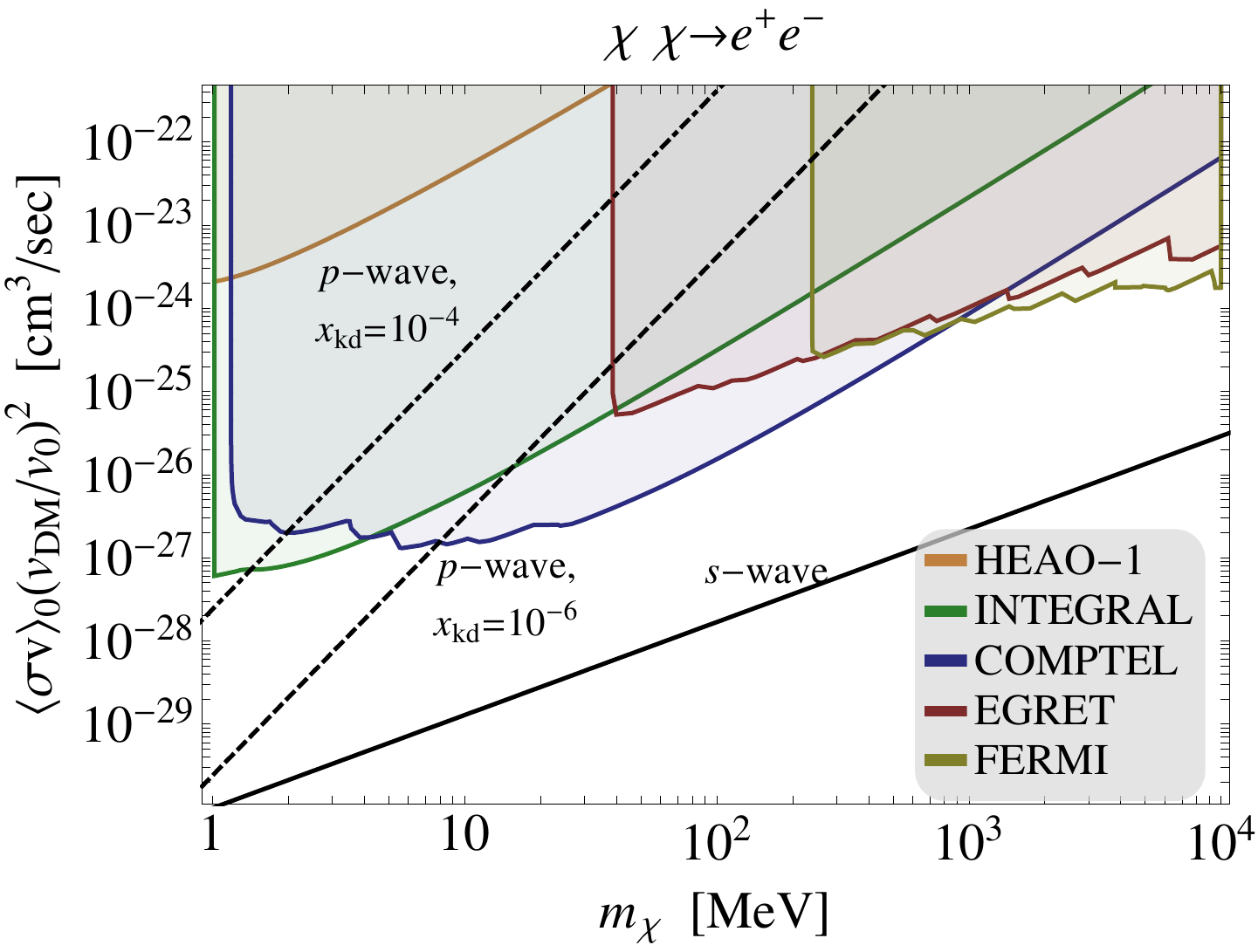}
\caption{ Bounds on the DM velocity-averaged annihilation cross-section $\langle \sigma v \rangle$ due to FSR off the process $\chi \chi \to e^+ e^-$. Regions as in \Fig{fig:darkphoton}.
Also shown is a comparison with the CMB constraint for DM annihilation that is $s$-wave (solid) or $p$-wave, the latter for 
two different kinetic-decoupling temperatures, $x_{\rm kd} \equiv T_{\gamma}/m_{\rm DM} = 10^{-4}$ (dash-dot) and $10^{-6}$ (dashed line), where we take $T_\gamma = 0.235$~eV at the CMB epoch (corresponding to $z_{\rm CMB} = 1000$). }
\label{fig:Ann-e+e-}
\end{center}
\end{figure}

Here we consider bounds on annihilating DM, specializing to the case of annihilation to $e^+ e^-$ (see also~\cite{Beacom:2004pe}).  
The differential photon spectrum for this case is
\begin{multline} \label{ann}
\frac{d N}{d E_\gamma} = \frac{2\alpha_{\rm EM}}{\pi E_\gamma} \frac1{ \pL 1 - \nu_e^2 \pR^{3/2} } \Bigg\{  \delta \pL 1-\nu_e^2 \pR  +\\  \bL 1-\lambda_\gamma+\half \lambda_\gamma^2 - \nu_e^2 \pL\frac32-\lambda_\gamma \pR + \half \nu_e^4\bR \ln \pL \frac{ 1- \lambda_\gamma - \delta}{1-\lambda_\gamma + \delta} \pR  \Bigg\},
\end{multline}
where we have defined $ \delta = \sqrt{\pL1-\lambda_\gamma \pR \pL 1-\lambda_\gamma - \nu_e^2 \pR }$.   The bounds are shown in \Fig{fig:Ann-e+e-}.   From \Tab{tab:jint}, we  see that these results are sensitive (within factors of a few) to the DM density profile (we use the NFW profile for all results), especially for experiments that observe regions near the center of the galaxy such as INTEGRAL and COMPTEL. 
For DM masses below $\sim 100 \mbox{ MeV}$ the bounds are stronger than the thermal annihilation cross-section around $3 \times 10^{-26} \mbox{ cm}^3/\mbox{s}$.

These bounds can be compared with those from CMB observations, which are very strong for $s$-wave processes.  Indeed, for DM masses below $\sim 7 \mbox{ GeV}$, the annihilation cross-section must be smaller than the thermal annihilation cross-section of $3 \times 10^{-26} \mbox{ cm}^3/\mbox{s}$.  At first sight, it appears that the diffuse photon bounds are not competitive with the CMB bounds.   However, $p$-wave annihilation rates may be larger in the galaxy today relative to the CMB epoch if the velocity of the DM at recombination is smaller than the galactic velocity which we take to be, $v_0 = 220$ km/sec.

The velocity of the DM at recombination depends on the kinetic decoupling temperature.  As long as the DM remains kinetically coupled to the plasma, its velocity is $v_{\rm DM} \sim \sqrt{3 T_\gamma/m_{\rm DM}}$.    
Once the DM kinetically decouples, however, it cools much more quickly: its temperature at redshift $z$ is $T_{\rm DM} = T_{\rm kd} \left(\frac{z}{z_{\rm kd}}\right)^2$, for a kinetic decoupling temperature $T_{\rm kd}$ at redshift $z_{\rm kd}$.  As a result, the DM velocity is 
\begin{eqnarray}
v_{\rm DM} & = & \sqrt{3 T_{\rm DM}/m_{\rm DM}} 
= \sqrt{3}x_\gamma \ x_{\rm kd}^{-1/2}\\ \nonumber 
& \simeq &2 \times 10^{-4} \left(\frac{T_\gamma}{1 \mbox{ eV}}\right)\left( \frac{1 \mbox{ MeV}}{m_{\rm DM}}\right)\left(\frac{10^{-4}}{x_{\rm kd}}\right)^{1/2},
\end{eqnarray} 
where we define $x_i \equiv T_i / m_{\rm DM}$.
The above  is easily smaller than the observed galactic velocity, even for very light DM.

We show in \Fig{fig:Ann-e+e-} the CMB constraint from $s$-wave processes, as well as the constraint from $p$-wave processes for $x_{\rm kd} = 10^{-4}$ and $~10^{-6}$, taking $T_\gamma = 0.235$ eV at the CMB epoch (corresponding to $z_{\rm CMB} = 1000$).  In order to compare the galactic and CMB constraints for both $s$- and $p$-wave annihilation, we show contours of $\langle \sigma v \rangle \propto \left( v_{\rm DM}/v_0\right)^{2(n -1)}$, where $n = 1(2)$ for $s(p)$-wave.  
We can see that the CMB constraints are always stronger than the diffuse photon constraints for $s$-wave annihilation.  
However, the diffuse constraints are stronger than the CMB constraints for $p$-wave annihilation, especially for larger 
kinetic-decoupling temperatures where the DM is colder.

\section{Conclusions and Future Improvements}
\label{sec:conclusions}

In this paper, we considered simplified models of DM with masses $\cO(\few \kev) \lesssim m_{\rm DM} \lesssim \cO(\few \gev)$, that can give rise to observable signals in X-ray and gamma-ray observatories via decays or annihilations. We found that bounds from HEAO-1, COMPTEL, INTEGRAL/SPI, EGRET and Fermi, even without dedicated searches, can already be very strong, even under conservative assumptions. 

For decaying light DM, constraints on the lifetime, $ \tau_{\rm DM}$, are in the range $10^{24} - 10^{28} \s$, where the weaker bounds typically apply in the case where DM decays to photons via FSR, while the stricter bounds apply when DM decays directly to photons.
On the other hand, for DM that annihilates to two electrically charged SM particles, we find that below a few hundred MeV the annihilation cross-section must be lower than the canonical thermal relic $s$-wave annihilation cross-section.  In this case, the existing CMB bounds are found to be stronger.   However, for $p$-wave suppressed annihilation, the CMB bounds become weaker than the diffuse constraints as the kinetic-decoupling temperature increases (and the DM at CMB becomes colder).

In addition to model-independent constraints, we also placed limits on  specific benchmark models of light DM:  hidden-photino DM, sterile-neutrino DM,  gravitino DM, dipole DM and hidden (pseudo-) scalar DM.   We found that the constraints from decaying DM are often stronger than other existing experimental, astrophysical, or cosmological constraints.  

We conclude that X-ray and gamma-ray observatories provide a powerful and independent probe of light DM.   To improve on the results presented here, dedicated searches are needed, where better background studies and optimized    regions in the sky are considered.   With the results above, the authors strongly encourage new studies in the hope of significantly widening the search window for dark matter.

\section*{Acknowledgements}

We  thank Kfir Blum, Germ\'an Arturo Gomez-Vargas, Shmuel Nussinov, Javier Redondo, Robert Shrock, and Andrew Strong for helpful discussions. RE is supported in part by the DoE Early Career research program DESC0008061 and by a Sloan Foundation Research Fellowship.  SDM and KZ are supported by the DoE under contract de-sc0007859, by NSF CAREER award PHY 1049896, and by NASA Astrophysics Theory grant NNX11AI17G. SDM is supported by the Fermilab Fellowship in Theoretical Physics. Fermilab is operated by Fermi Research Alliance, LLC, under Contract No.~DE-AC02-07CH11359 with the United States Department of Energy.  TV is supported in part by a grant from the Israel Science Foundation, the US-Israel Binational Science Foundation,  the EU-FP7 Marie Curie, CIG fellowship and  by the I-CORE Program of the Planning and Budgeting Committee and The Israel Science Foundation (grant NO 1937/12).

\appendix

\section{Constraints with fits to astrophysical backgrounds}\label{sec:gof}

In this paper, we derived robust, conservative constraints by only taking into account the DM signal, as described in Sec.~\ref{sec:stat}.  Stronger constraints can be obtained by fitting the DM signal simultaneously with the different astrophysical background components.  This could improve the constraints especially if the DM signal spectrum has a sharp feature like a line or an edge (as appears in an FSR spectrum). However, for softer spectra, while the constraints may be formally stronger, they also suffer from larger systematic uncertainties, since the background components are not known precisely.  Furthermore, the isotropic extragalactic flux, which contributes an $\mathcal{O}(1)$ amount to the diffuse galactic signal at high galactic latitudes, can smear out any spectral shapes~\cite{Hensley:2012xj}. 

\begin{figure}[t]
\begin{center}
\includegraphics[width=0.8\textwidth]{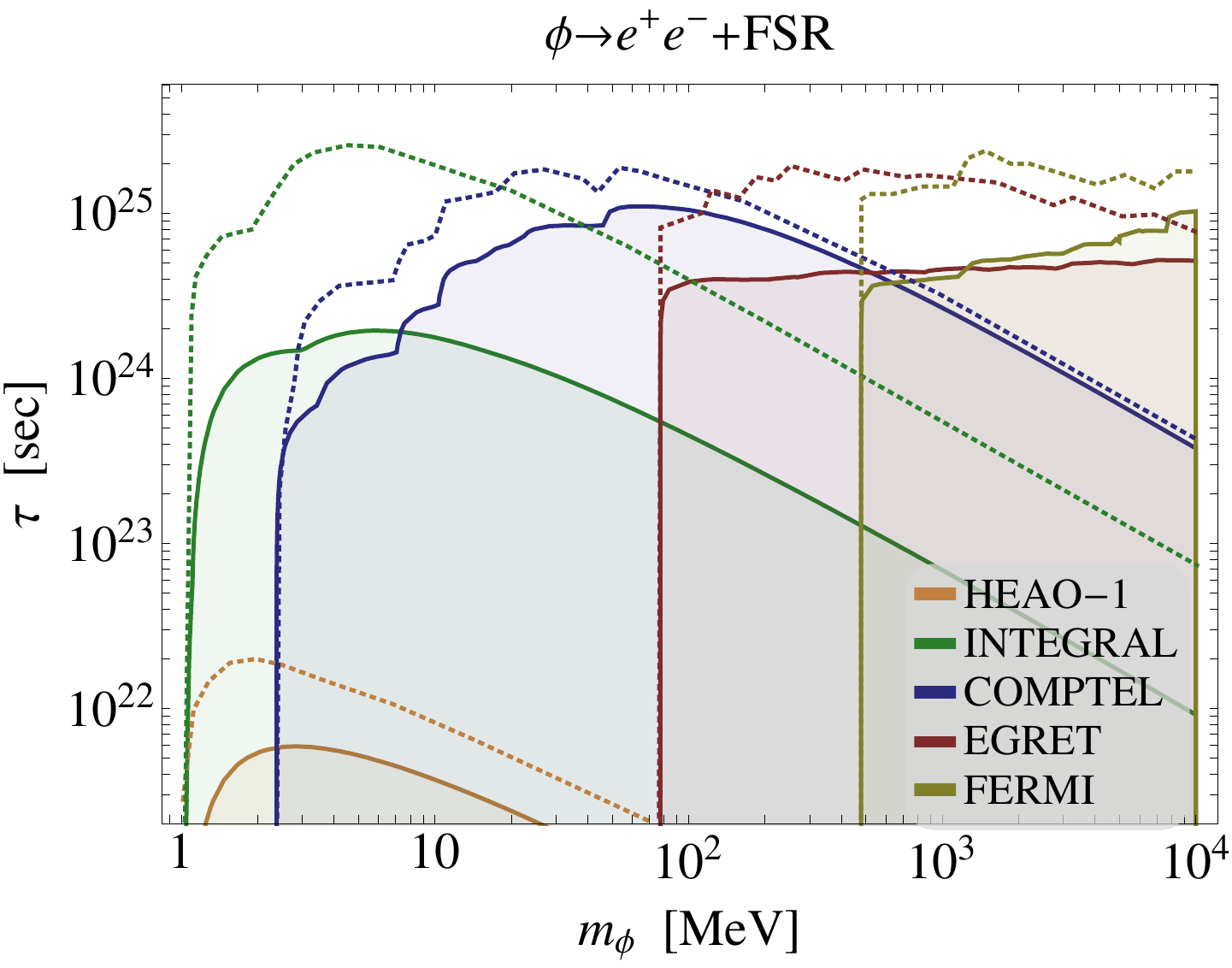}
\caption{Comparison of signal-only constraint (solid) and a $\chi^2$ goodness-of-fit test (dotted) for each experiment taking the sample spectrum from scalar DM decay to $e^+e^-$ pairs that emit FSR.  We show the limits derived from the data described in Sec.~\ref{sec:Indirect}: HEAO-1 (orange), INTEGRAL (green), COMPTEL (blue), EGRET (red), and Fermi (yellow). }
\label{stat-compare}
\end{center}
\end{figure}

To illustrate the improvements possible with using a simultaneous fit of signal and backgrounds, we use the background components as derived by the different collaborations in~\cite{Gruber:1999yr,Bouchet:2008rp,COMPTEL,Strong:2003ey,FermiLAT:2012aa} and perform a
na\"ive $\chi^2$ goodness-of-fit test (GOF) in \Fig{stat-compare}. For the GOF, we take as many distinct background components as have been identified by each collaboration, and, keeping the slopes fixed, allow the normalizations to float. At each point in the $\tau-m_{\rm DM}$ plane, we add the putative DM signal and minimize the $\chi^2$ of the background plus the signal with respect to the free normalization parameters. For the HEAO-1 backgrounds, we minimize the $\chi^2(S+B)$ by allowing the overall normalization of the broken power law suggested by the collaboration to float. 
For the INTEGRAL backgrounds, we allow the normalizations of the three smooth background components identified by the collaboration to float independently, and again minimize the $\chi^2(S+B)$. These components are a power law with a spectral index $n_s=1.55$, a curved component that is the exponential tail of a flat power law (with cutoff around 7.5 keV), and the smooth diffuse component from extragalactic $e^+e^-$ annihilation. The COMPTEL collaboration identifies a single smooth background component with index $n_s=2.4$, and again we minimize over the normalization of this background. The EGRET and Fermi data are dominated by the systematic error on the effective area, so we take the total shapes as given by the collaborations and allow the normalizations on the entire background shape to float simultaneously.  We show the comparison in \Fig{stat-compare}, and we find that the GOF improves the constraints, but only by at most an order of magnitude.

\end{document}